\documentclass[aps,pre,longbibliography]{revtex4-1}
\usepackage[dvips]{graphicx}
\usepackage{color}

\usepackage{morefloats}
\usepackage{subfigure}

\begin{document}
\newcommand{\vphi}{\varphi}
\newcommand{\bq}{\begin{equation}}
\newcommand{\ba}{\begin{eqnarray}}
\newcommand{\eq}{\end{equation}}
\newcommand{\ee}{\end{equation}}
\newcommand{\ea}{\end{eqnarray}}
\newcommand{\tchi} {{\tilde \chi}}
\newcommand{\tA} {{\tilde A}}
\newcommand{\sech} { {\rm sech}}
\newcommand{\pstar}{\mbox{$\psi^{\ast}$}}
\newcommand {\bPsi}{{\bar \Psi}}
\newcommand {\bpsi}{{\bar \psi}}
\newcommand {\tu} {{\tilde u}}
\newcommand {\tv} {{\tilde v}}
\newcommand{\dq}{{\dot q}}

\title{Speed-of-light pulses in the massless nonlinear Dirac equation with a potential}

\author{Niurka R.\ Quintero}
\email{niurka@us.es}
\affiliation{Instituto Carlos I de F\'\i sica Te\'orica y Computacional, Universidad de Granada, 18071 Granada, Spain}
\author{Franz G.  Mertens}
\email{franzgmertens@gmail.com}
\affiliation{Physikalisches Institut, Universit\"at Bayreuth, D-95440 Bayreuth, Germany} 
\author{Fred Cooper}
\email{cooper@santafe.edu}
\affiliation{Santa Fe Institute, Santa Fe, NM 87501, USA}
\affiliation{Theoretical Division and Center for Nonlinear Studies, 
Los Alamos National Laboratory, Los Alamos, New Mexico 87545, USA}
\author{Avadh Saxena}
\email{avadh@lanl.gov}
\affiliation{Theoretical Division and Center for Nonlinear Studies, 
Los Alamos National Laboratory, Los Alamos, New Mexico 87545, USA}
\author{A. R. Bishop} 
\email{arb@lanl.gov} 
\affiliation
{Los Alamos National Laboratory,  Los Alamos, NM 87545, USA}
\date{\today}
\begin{abstract}
We consider the massless nonlinear Dirac (NLD) equation  in $1+1$ dimension with scalar-scalar self-interaction $
\frac{g^2}{2}  (\bar{\Psi} \Psi)^2$ in the presence of three external electromagnetic potentials  $V(x)$, a potential barrier, a constant potential, and a potential well.  
 By solving numerically the NLD equation, we find that, for all three cases, after a short transit time, the initial pulse breaks into two pulses which are solutions of the 
\textit{massless linear} Dirac equation traveling in opposite directions with the speed of light. During this splitting the charge and the energy are conserved, whereas the momentum is conserved when the solutions possess specific  symmetries. For the case of the constant potential, we derive exact analytical solutions of the massless NLD equation that are also solutions of the massless linearized Dirac equation.   
\end{abstract}
\maketitle

\section{Introduction} \label{sec1}

The relativistic generalization of the nonlinear Schr\"odinger (NLS) equation, namely the nonlinear Dirac (NLD) equation  
has emerged as a natural model in many physical systems, such as extended particles \cite{finkelstein:1956,heisenberg:1957}, 
light solitons in waveguide arrays and experimental realization of an optical analog for relativistic quantum mechanics  \cite{tran:2014},
Bose-Einstein condensates in honeycomb optical lattices \cite{haddad:2009,haddad:2015,arevalo:2016}, and 
phenomenological models of quantum chromodynamics
\cite{fillion:2013}, among many others. 

 Solitary waves in the $1+1$  dimensional nonlinear Dirac (NLD) equation  have been
studied \cite{lee:1975,nogami:1992} in the past in the case of massive Gross-Neveu \cite{gross:1974} (with $N = 1$, i.e. just one localized fermion) and 
massive Thirring \cite{thirring:1958} models). Soler \cite{soler:1970} proposed in 1970 that the self-interacting $4$-Fermi theory was a useful model to study extended fermions. Subsequently,
Strauss and Vázquez \cite{strauss:1986} were able to study the stability of this model under dilatation and found the domain of
stability for the Soler solutions. These solutions are solitary waves which can have either one- or two humps, depending on the value of the frequency $\omega \in (0,1)$. Recent studies using a split operator method suggested that all stable NLD solitary waves have a one-hump profile, but not all one-hump waves are stable, while all waves with two humps are unstable \cite{shao:2014}.  
In particular for the scalar-scalar self-interaction $(g^2/2) (\bar{\Psi} \Psi)^2$ the
solitary waves were stable in simulations only if $\omega \in [0.56,1)$ \cite{shao:2014}.

The interaction between solitary waves of different
initial charge was studied in detail for the scalar-scalar  case in the work of Alvarez and Carreras \cite{alvarez:1981} by Lorentz boosting the
static solutions and allowing them to scatter. 
More accurate simulations have been performed by Shao and Tang  in \cite{shao:2005}, where a new quasi-stable long-lived oscillating bound state from the binary collisions of a single-humped soliton and a two-humped soliton was observed. The dynamics of single solitons also has been studied for the NLD 
equation with external electromagnetic fields as well as under forcing conditions \cite{nogami:1992,mertens:2012,mertens:2016}. The functional shape of the initial soliton does not change, however the soliton is accelerated in the ramp potential, or its center oscillates around the initial position in the cases of harmonic and periodic potentials. 
For smaller values of $\omega$  the soliton,  after some transient time, is again unstable.    

 In this paper we 
study the very interesting behavior of the time evolution of  solitary waves solutions of the Soler model placed at the origin when we add a potential barrier, a constant potential and a potential well,  where the evolution is governed by the  massless NLD equation in that external potential. For that problem we will numerically show that the initial pulse, regardless of the value of $\omega \in (0,1)$, rapidly ``decays" into two pulses moving in opposite directions and  preserving the charge $Q$ as well as the energy $E$ of the initial pulse. We find that the self-interaction goes to zero after a critical time 
and one is left with two solutions of the massless linear Dirac equation having solitary wave shape, traveling at the speed of light in
opposite directions.  For the case of a constant external potential we find exact solutions to the full massless NLD equation which are also exact solutions of the massless linear Dirac equation.  These analytic solutions provide insight into our numerical results as they represent a single pulse at time zero which becomes two pulses moving in opposite directions at later times. 

This paper is organized as follows: In Sec. II, we present the numerical solutions of the massless NLD equation for: (a) a potential barrier, (b) a constant potential and (c) a potential well. 
In Sec. III we find the exact analytical solutions of the massless NLD equation with a constant external potential. 
In Sec. IV we discuss our main findings and conclusions. The conservation of the charge and the energy is discussed  in the Appendix, where we also show that if the external potential is symmetric, $V(x)=V(-x)$, for certain symmetries of the massless NLD equation the momentum is also conserved. 

\section{Numerical solutions for the massless NLD equation} \label{sec2}

The NLD equation in $1+1$ dimensions with scalar-scalar self-interaction is given by
\begin{equation} \label{eq1}
i \gamma^{\mu} \partial_{\mu} \Psi - m \Psi + g^2 (\bar{\Psi} \Psi) \Psi=0, 
\end{equation}
where 
\begin{equation} \label{eq2}
\Psi(x,t)=\left(  
\begin{array} {cc}
     \psi(x,t) \\
     \chi(x,t) \\ 
   \end{array} \right),
\end{equation}
is a two-component  spinor field,  $\bar{\Psi}=\Psi^{\dag} \gamma^{0}$ is the adjoint spinor, $\gamma^{\mu}$ are the Dirac  matrices.  We choose the representation 
$\gamma^0 = \sigma_3$ and $\gamma^1= i \sigma_2$, where $\sigma_j$ are the Pauli matrices, $m$ is the mass and $g$ is the coupling constant. 

We add electromagnetic interactions through the gauge covariant derivative
\begin{equation} \label{eq3}
i \partial_{\mu} \Psi \rightarrow (i \partial_{\mu}-e A_{\mu}) \Psi.
\end{equation}
Using the freedom of gauge invariance, we choose the axial gauge $A_1=0$, $e A_0=V(x)$. In this gauge the nonlinear Dirac equation becomes
\begin{equation} \label{eq4}
i \gamma^{\mu} \partial_{\mu} \Psi - m \Psi + g^2 (\bar{\Psi} \Psi) \Psi=\gamma^0 V(x) \Psi.  
\end{equation}
As initial condition (IC) for the numerical solution of Eq.\ (\ref{eq4}) we take the exact static solitary wave solution of Eq.\ (\ref{eq1}) 
\cite{cooper:2010,mertens:2012}
\begin{equation} \label{eq5}
\psi(x,0)=A(x), \quad \chi(x,0)=i B(x),   
\end{equation}
with 
\begin{eqnarray}\label{eq6}
A(x)&=&\frac{\sqrt{2} \beta_1 \sqrt{m+\omega}}{g} \frac{\cosh \beta_1 x}{m+\omega \cosh 2 \beta_1 x}, \\ \label{eq7}
B(x)&=& \frac{\sqrt{2} \beta_1 \sqrt{m-\omega}}{g} \frac{\sinh \beta_1 x}{m+\omega \cosh 2 \beta_1 x},  
\end{eqnarray}
where $\beta_1=\sqrt{m^2-\omega^2}$.  We cannot set $m=0$, because $\beta_1$ would be imaginary. Therefore we replace $m$ in Eqs.\ (\ref{eq6})-(\ref{eq7}) by a new parameter $\mu$.  That is, we use as IC Eq.\ (\ref{eq5}) with 
\begin{eqnarray}\label{eq8}
A(x)&=&\frac{\sqrt{2} \beta \sqrt{\mu+\omega}}{g} \frac{\cosh \beta x}{\mu+\omega \cosh 2 \beta x}, \\ \label{eq9}
B(x)&=& \frac{\sqrt{2} \beta \sqrt{\mu-\omega}}{g} \frac{\sinh \beta x}{\mu+\omega \cosh 2 \beta x},  
\end{eqnarray}
where $\beta=\sqrt{\mu^2-\omega^2}$. This IC has essentially the same properties as Eqs.\ (\ref{eq6})-(\ref{eq7}): it is localized, $A(x)$ is symmetric and $B(x)$ is antisymmetric. 

As we want to investigate the massless NLD equation, we now set $m=0$ in Eq.\ (\ref{eq4}). Since the fields can be scaled, without loss  of generality we can consider $g=1$. Moreover, 
we choose $\mu=1$ which means that $\omega$ is in the range $0< \omega<1$. 

For the potential in Eq.\ (\ref{eq4}) we consider three cases: a potential barrier $V_{1}(x)=\omega + \mu\, \sech\, 2 \beta x$, a constant potential $V_{2}(x)=\omega$, and a potential well $V_{3}(x)=\omega - \mu \, \sech\, 2 \beta x$. 
For all three cases we have included the constant term $\omega$, because it will turn out 
that the energy, $E= \int \, dx \, T^{00}$, in the second case is equal to $\omega Q$. The energy and the charge are conserved (see the Appendix).  Here $T^{00}$ denotes the density of the energy. 

Our numerical simulations have been performed by means of a fourth-order Runge-Kutta method. We choose $N+1$ points starting at $n=0$ and vanishing boundary conditions 
$\Psi(\pm L, t)=0$. The other parameters related with the discretization of the system are $x \in [-L,L]$, $\Delta x=0.02$, $L=100$ and $\Delta t=0.0001$. 
  
\begin{figure}[ht!]
\begin{center}
\begin{tabular}{cc}
 \includegraphics[width=7.0cm]{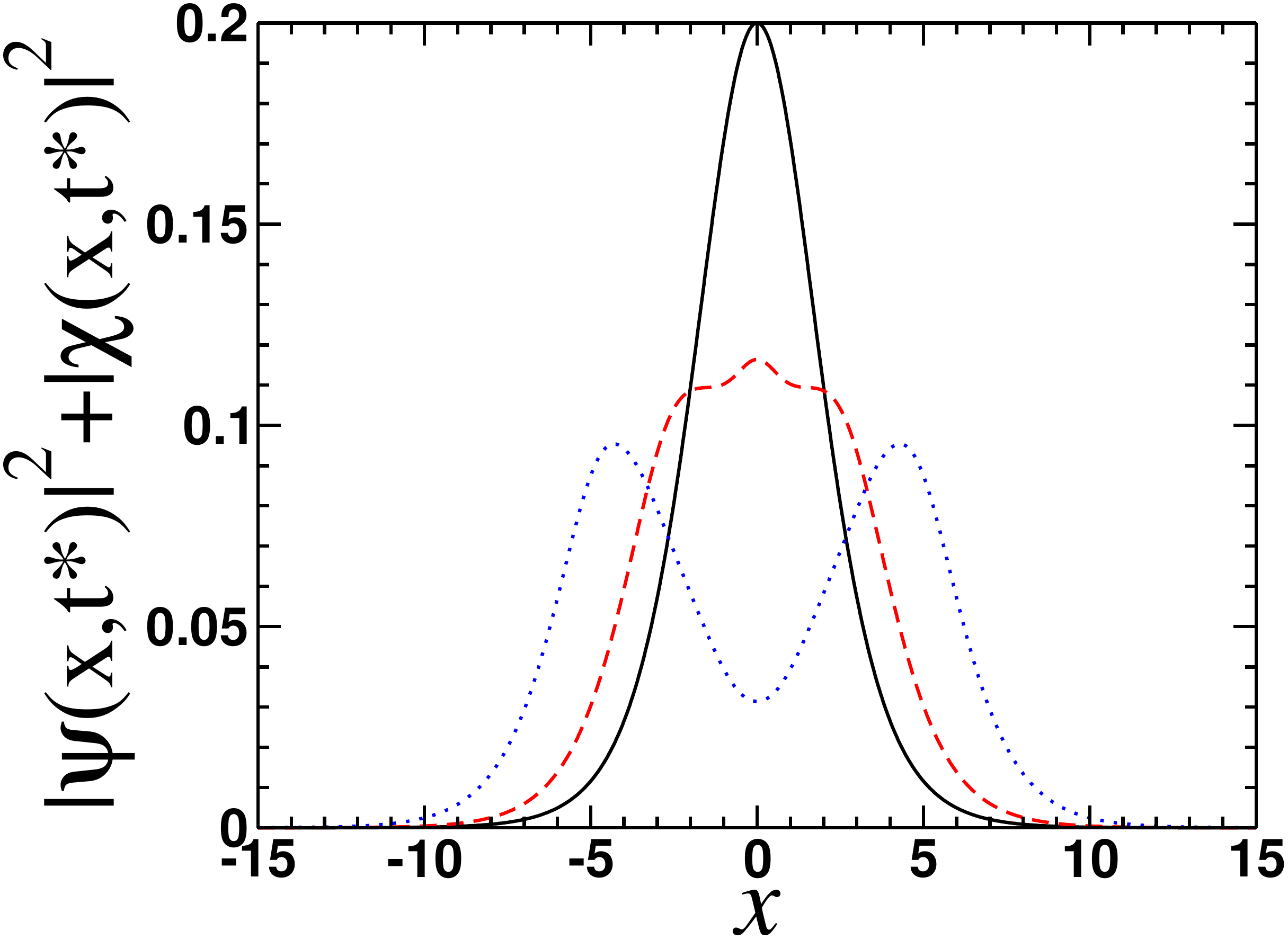} &   
\includegraphics[width=7.0cm]{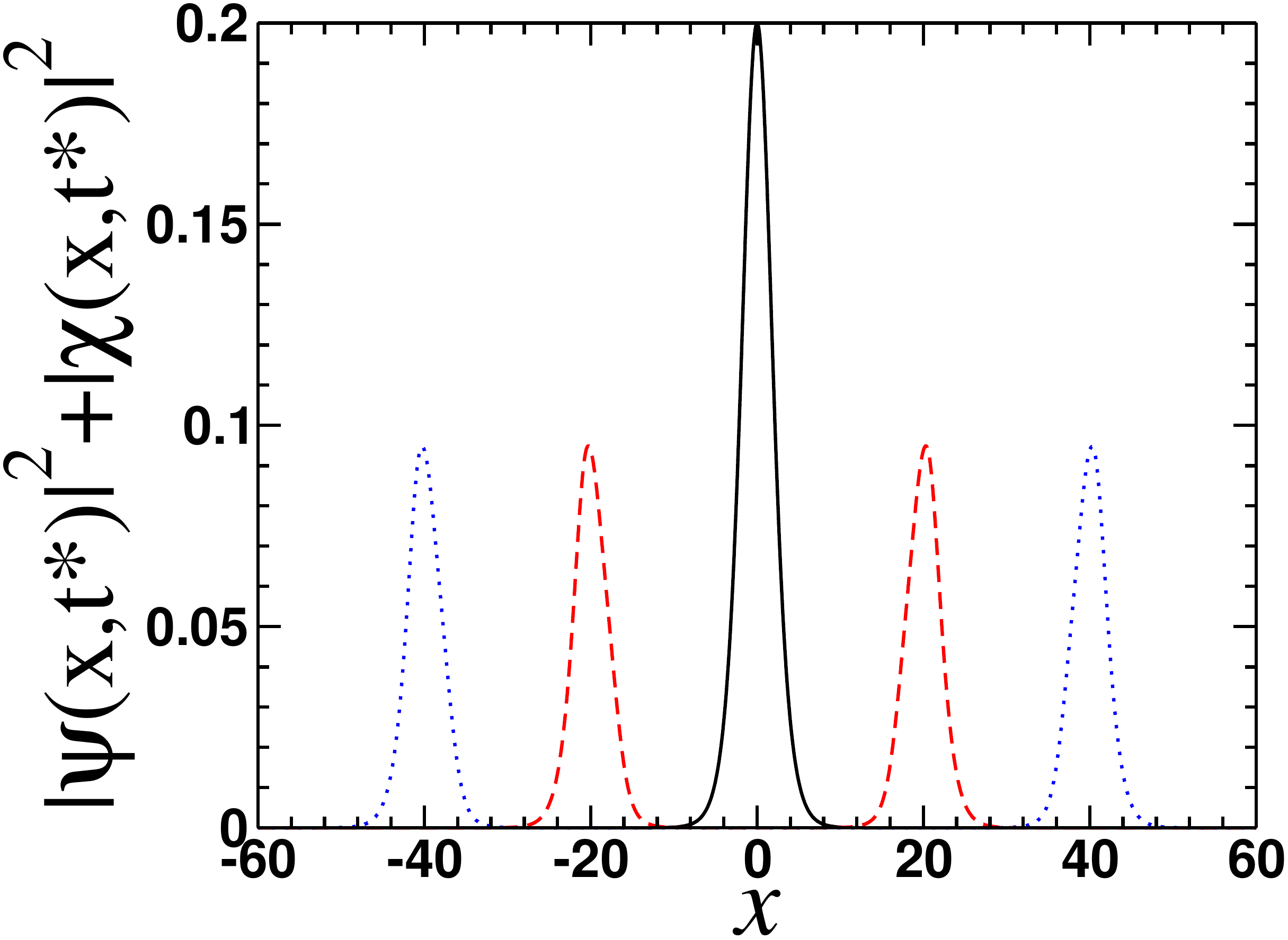} \\
\includegraphics[width=7.0cm]{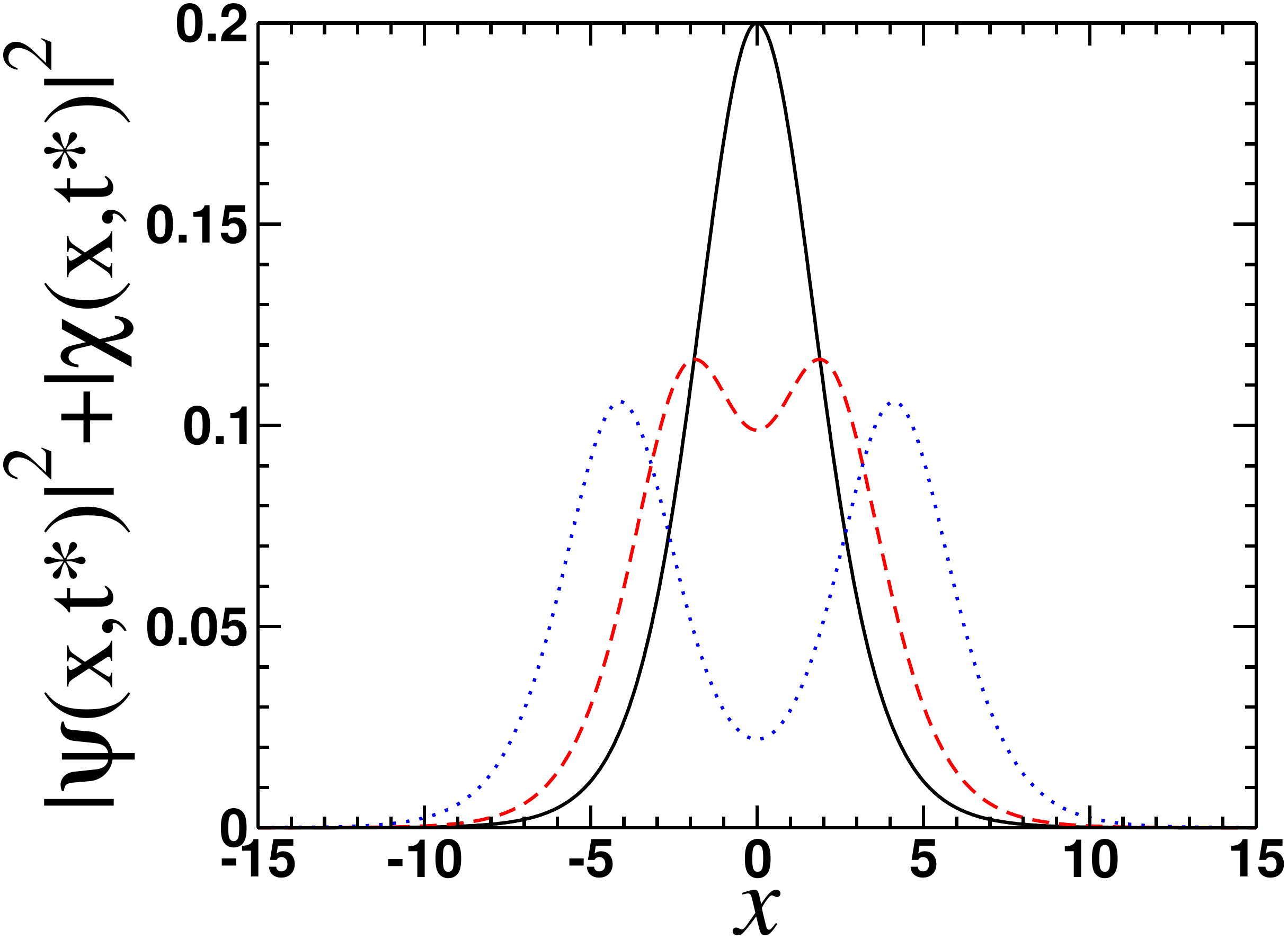} &  
 \includegraphics[width=7.0cm]{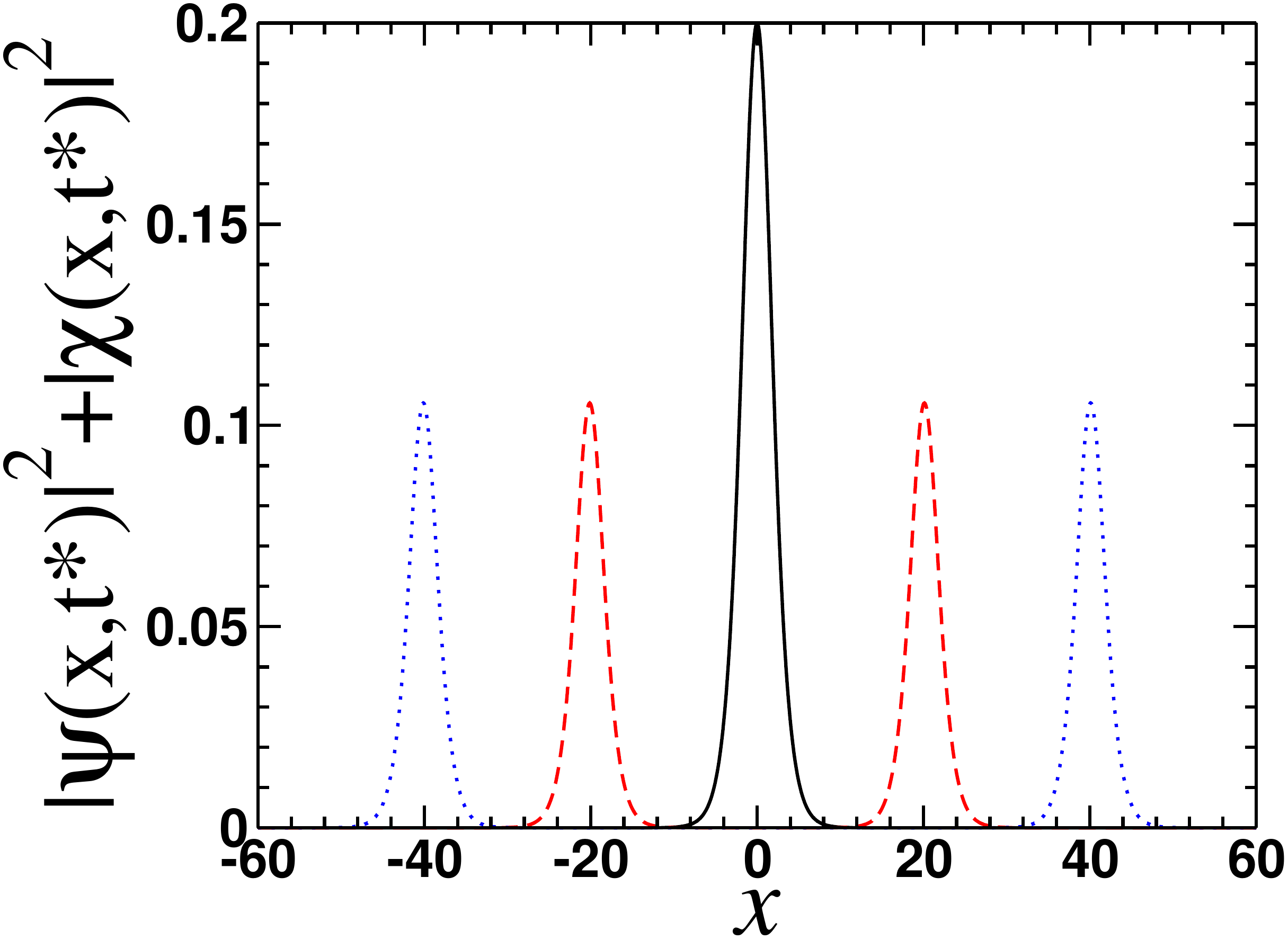} \\
   \includegraphics[width=7.0cm]{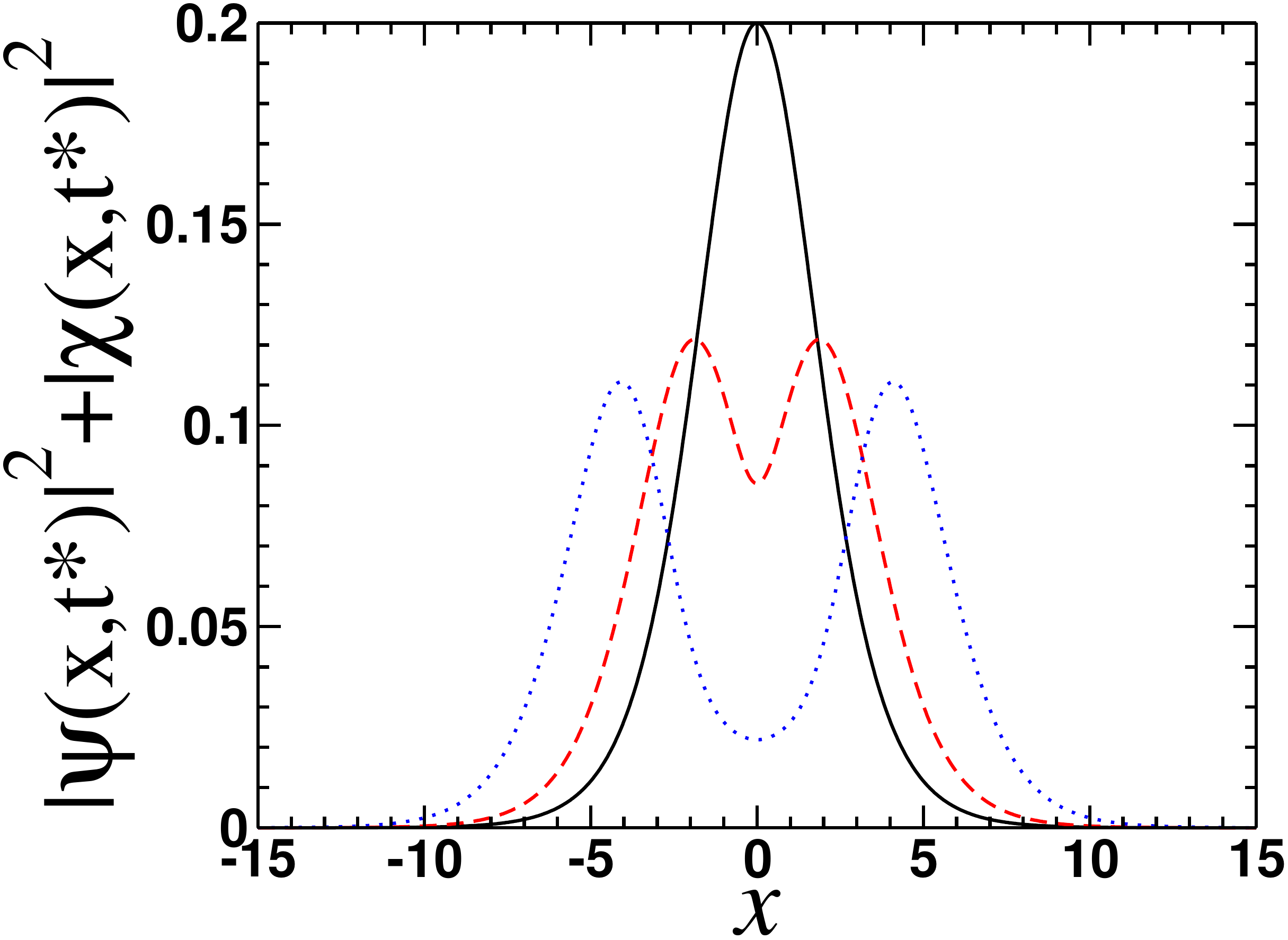} &   
 \includegraphics[width=7.0cm]{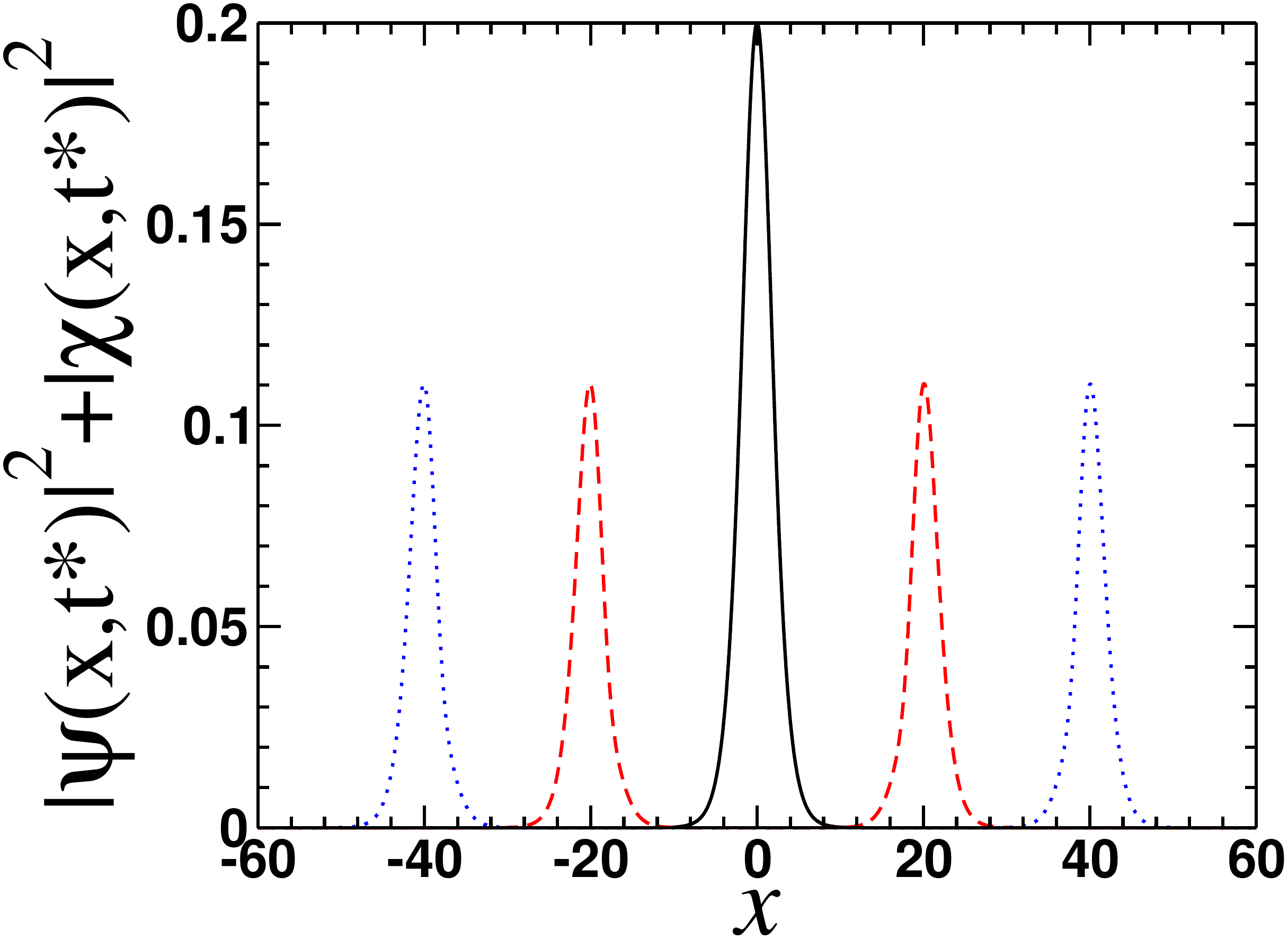}
\end{tabular}
\end{center}
\caption{Snapshots of the charge density. Left side: $t^{\star}=0$ (black solid lines), $t^{\star}=2$ (red dashed lines) and $t^{\star}=4$ (blue dotted lines). Right side:  $t^{\star}=0$ (black solid lines), $t^{\star}=20$ (red dashed lines) and $t^{\star}=40$ (blue  dotted lines). Upper panels: potential barrier 
$V_{1}(x)=\omega + \mu\, \sech\, 2 \beta x$, middle panels: constant potential $V_{2}(x)=\omega$, lower panels: potential well $V_{3}(x)=\omega - \mu \, \sech\, 2 \beta x$. 
Parameters: $g=1$,  $\omega=0.9$ and $\mu=1$. IC: 
Eqs.\ (\ref{eq5}) with (\ref{eq8})-(\ref{eq9}).    
 }
\label{fig1} 
\end{figure}

We first choose the parameter $\omega=0.9$ for which the initial charge density   
\begin{equation} \label{eq10}
\rho(x,0)=|\psi(x,0)|^2+ |\chi(x,0)|^2=A(x)^2+B(x)^2,   
\end{equation}  
is a single pulse. The simulation shows that this initial pulse splits into two pulses which move in opposite directions with the speed of light (Fig.\ \ref{fig1}). 
Interestingly, this scenario is the same for the three potentials used for Fig.\ \ref{fig1}.   
The splittings take place for short times $0<t<t_s \approx 10$ and differ for the different potentials (Fig.\ \ref{fig1}). 

\begin{figure}[ht!]
\begin{center}
\begin{tabular}{cc}
\includegraphics[width=7.0cm]{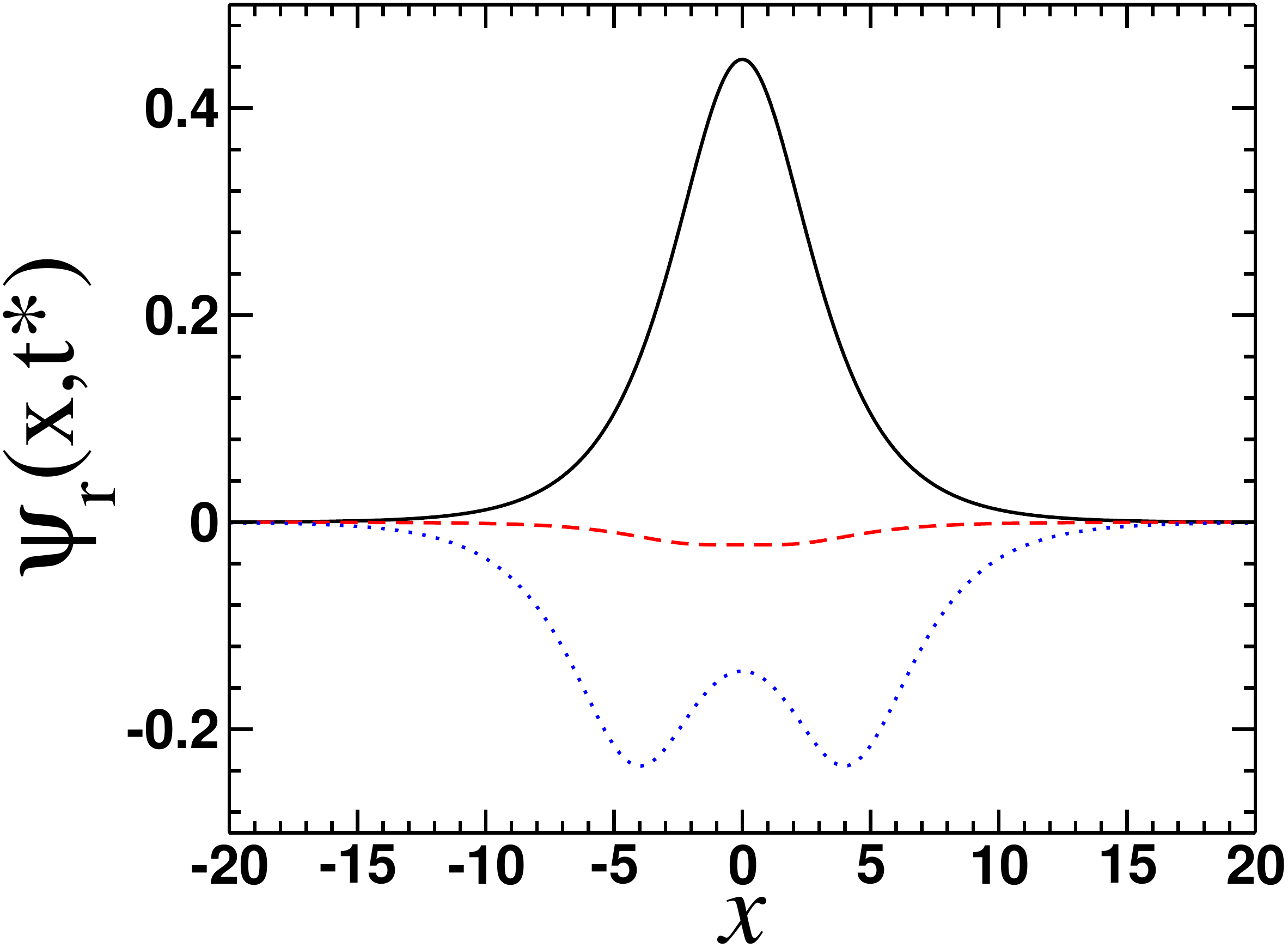} & 
  \includegraphics[width=7.0cm]{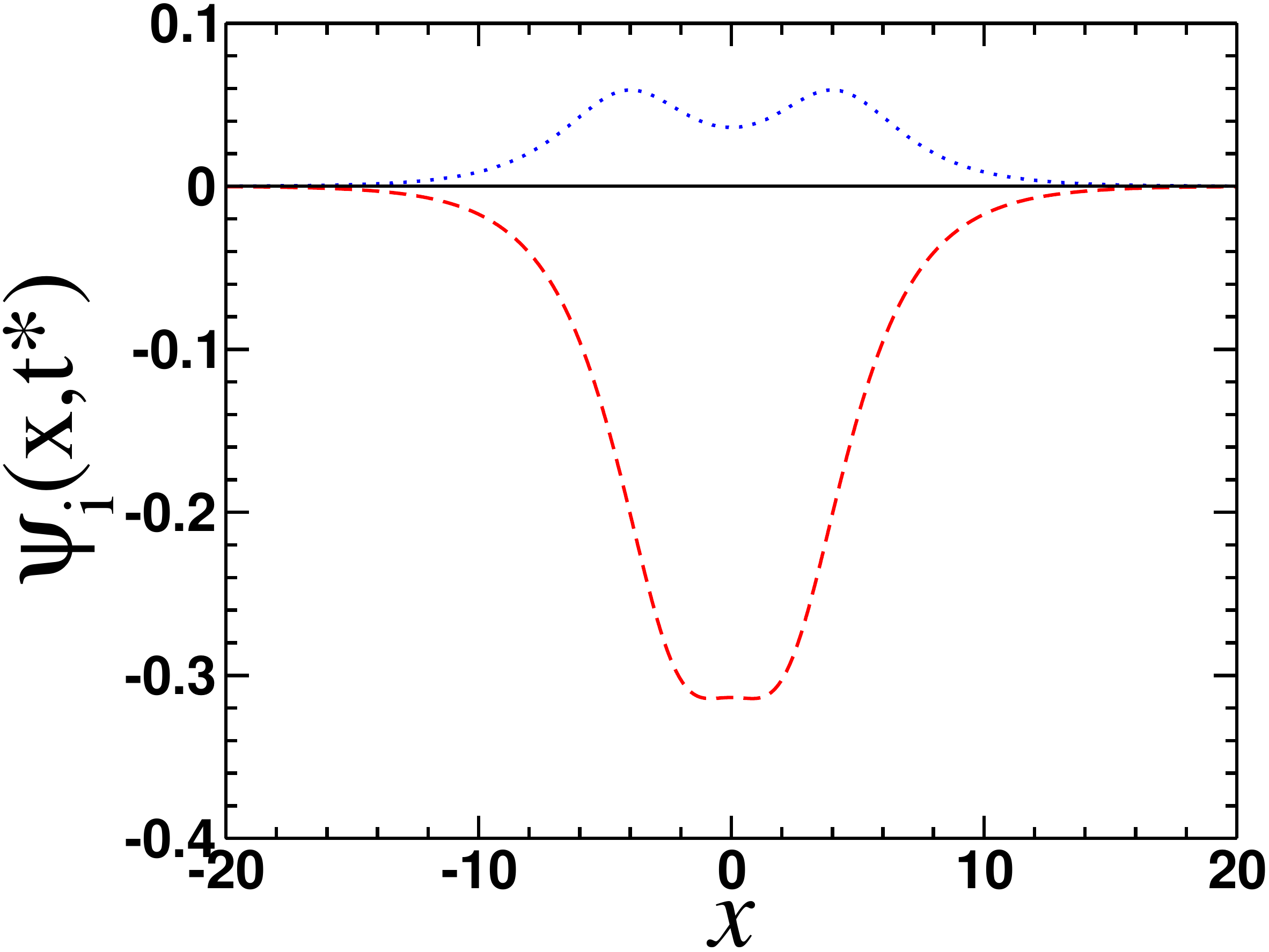} \\
 \includegraphics[width=7.0cm]{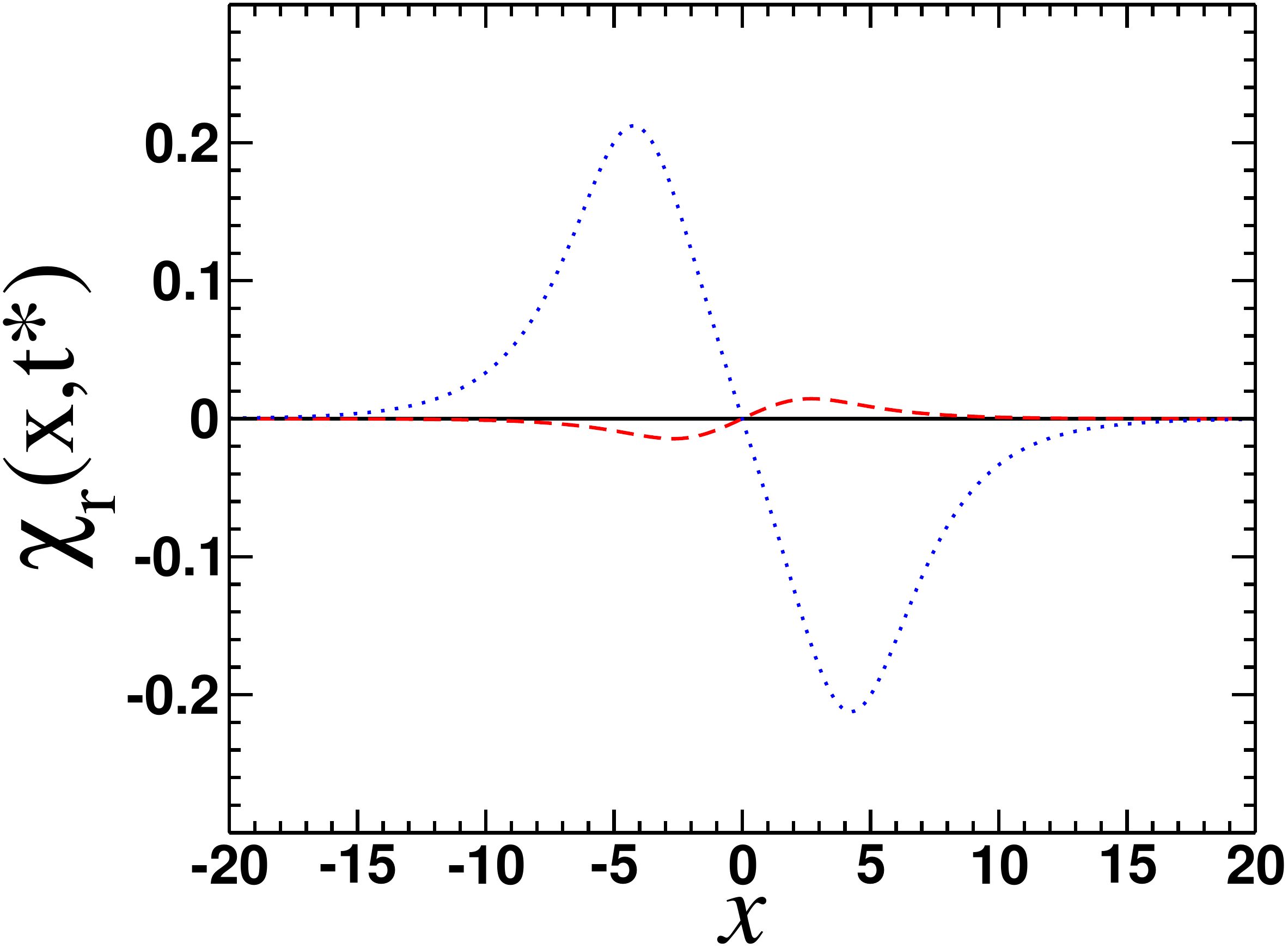} &   
 \includegraphics[width=7.0cm]{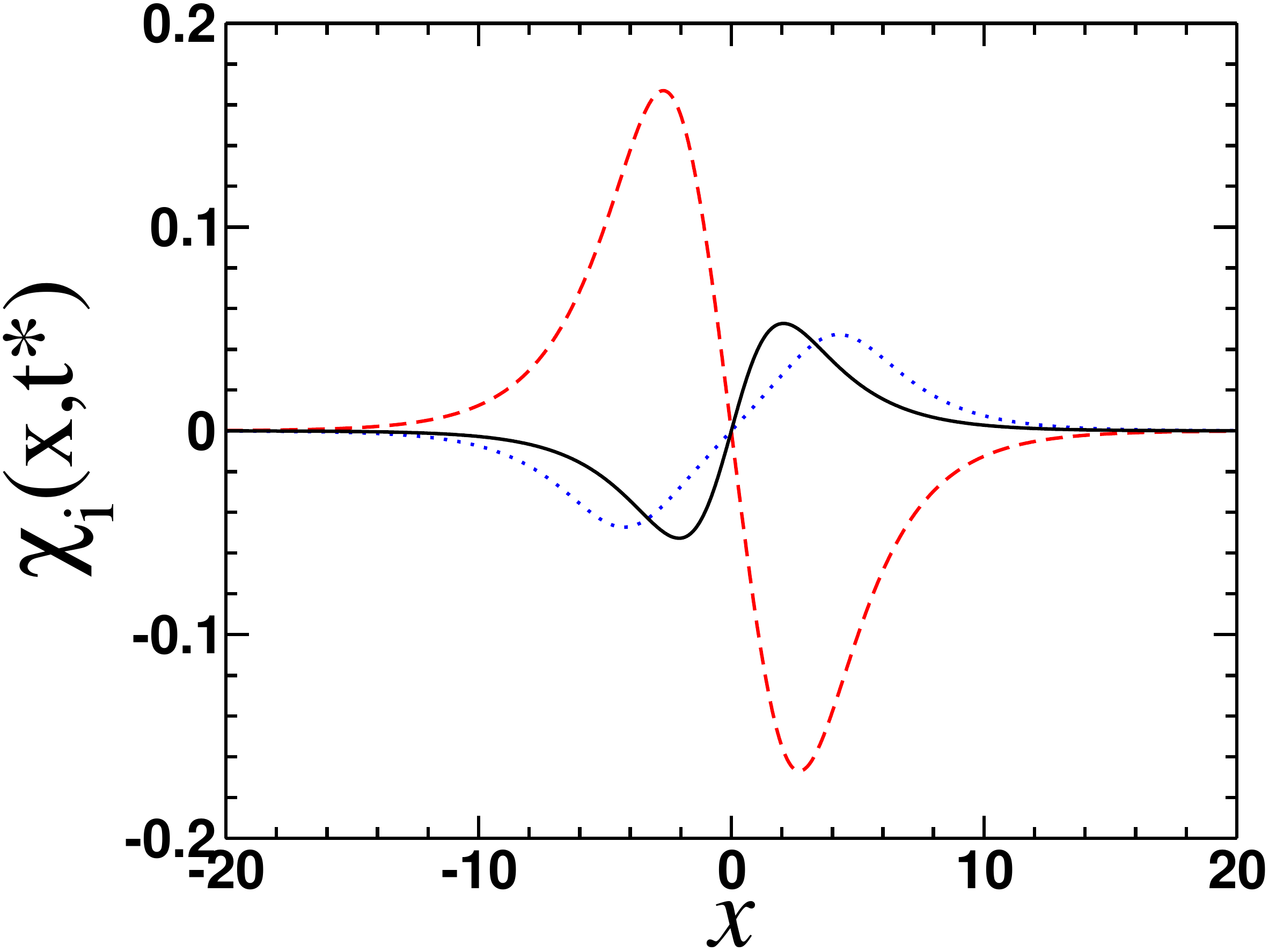}  
\end{tabular}
\end{center}
\caption{  
Snapshots of the real and imaginary parts of the spinor components $\psi$ 
and $\chi$.  Black solid lines: $t^{\star}=0$, red dashed lines: $t^{\star}=2$, blue dotted lines: $t^{\star}=4$. Constant potential $V_2(x)=\omega$. Parameters and ICs: 
same as in Fig.\ \ref{fig1}.
 }
\label{fig7} 
\end{figure}

\begin{figure}[ht!]
\begin{center}
\begin{tabular}{cc}
\includegraphics[width=7.0cm]{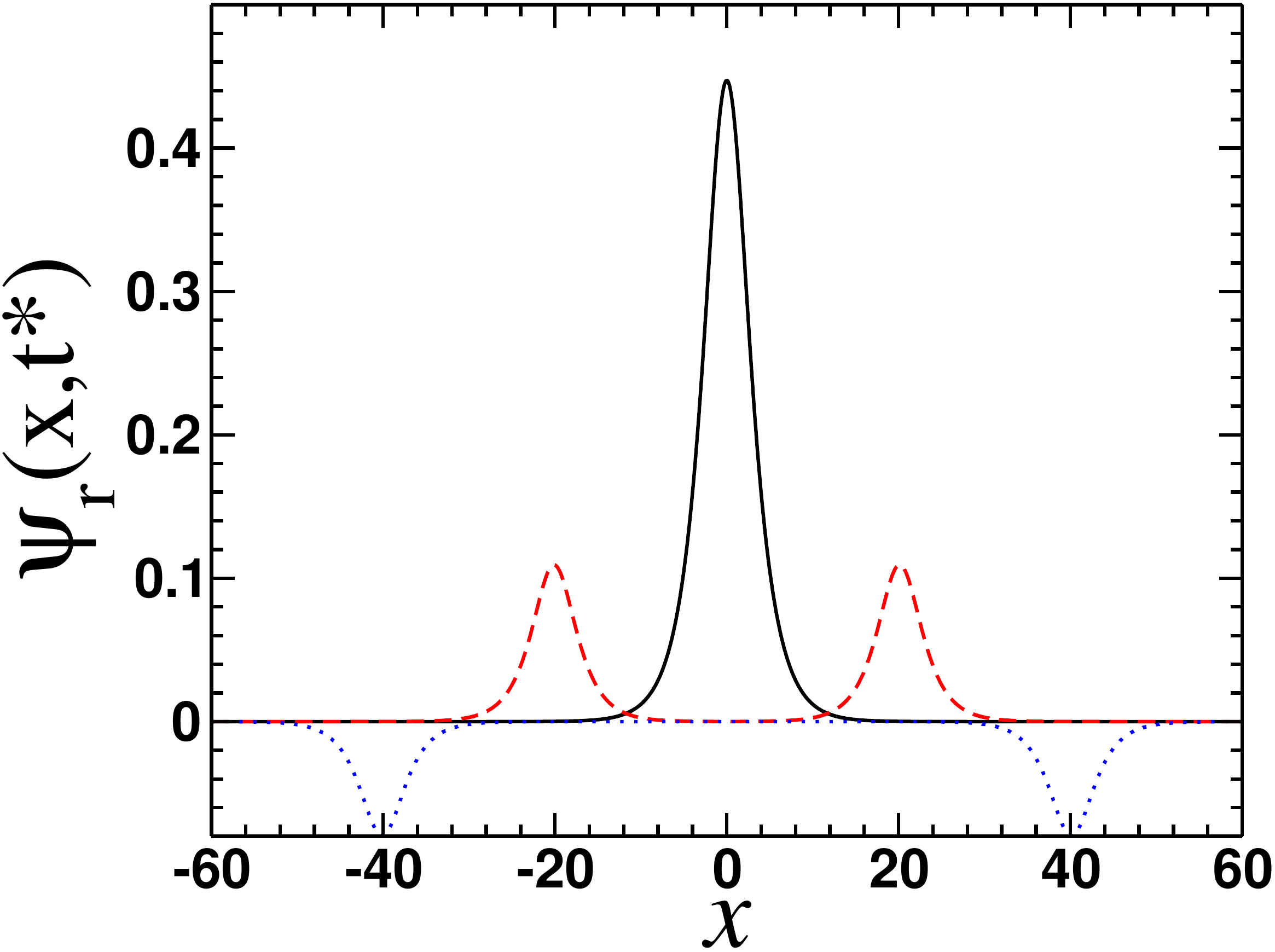} & 
  \includegraphics[width=7.0cm]{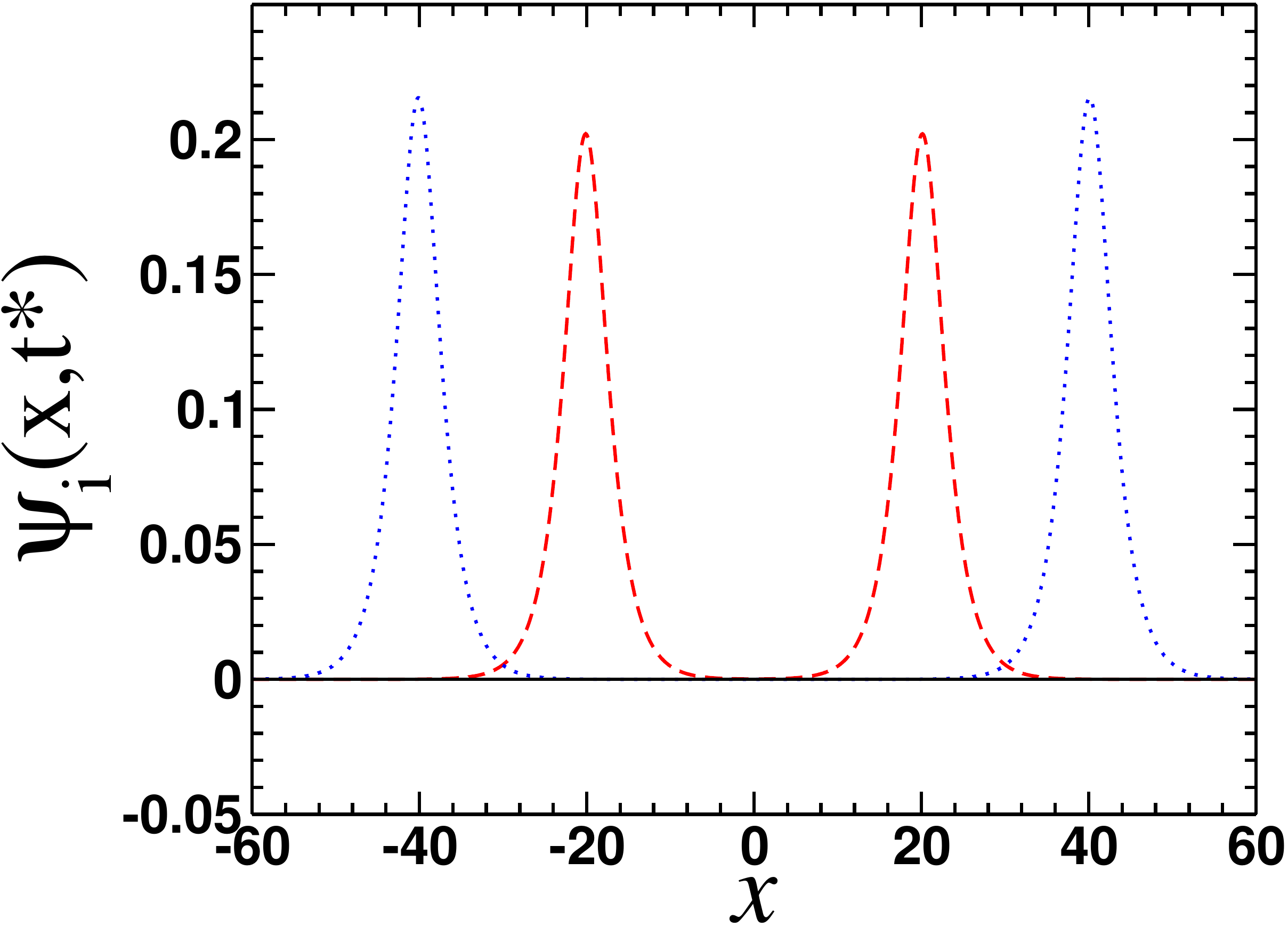} \\
 \includegraphics[width=7.0cm]{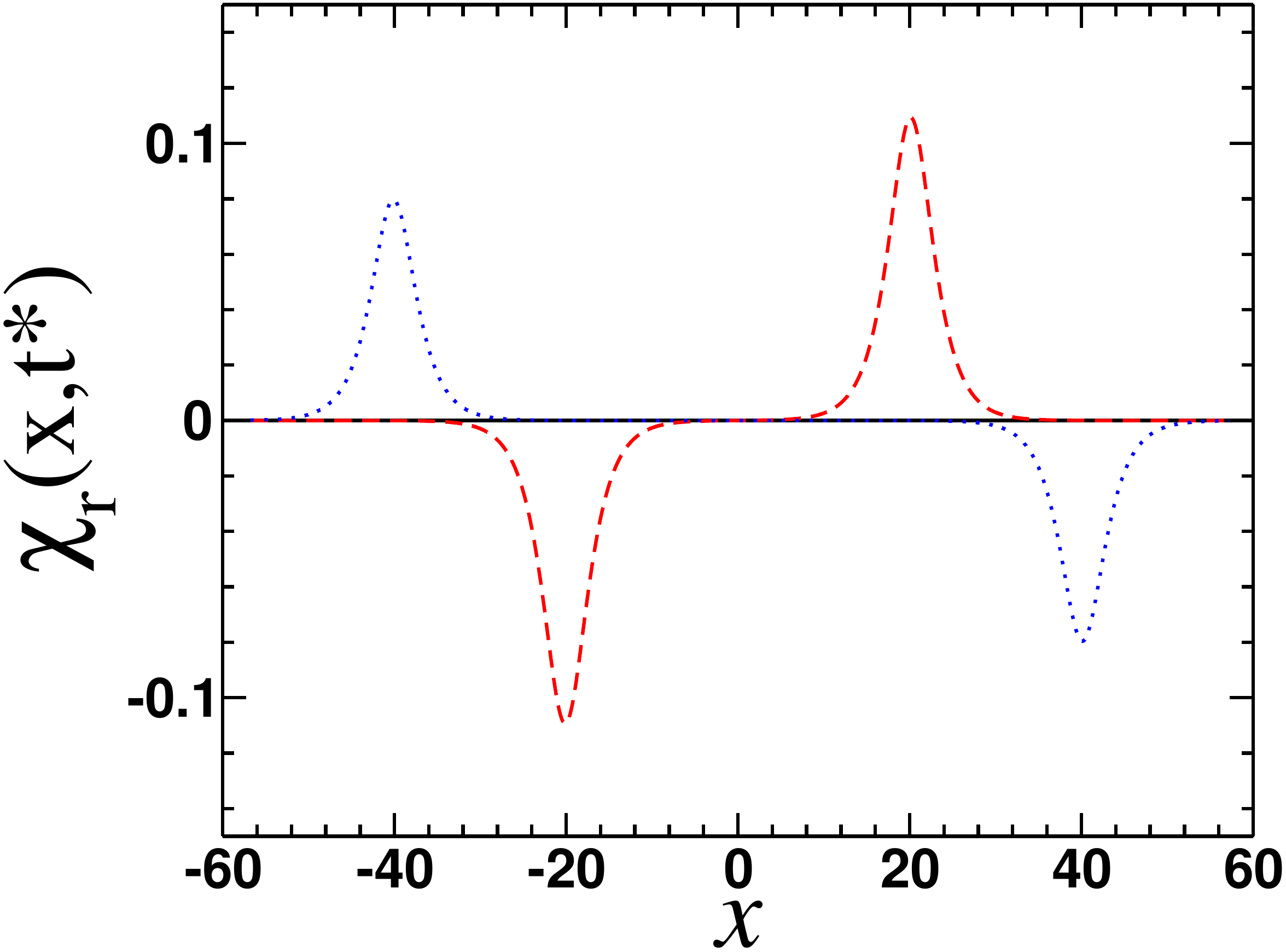} &   
 \includegraphics[width=7.0cm]{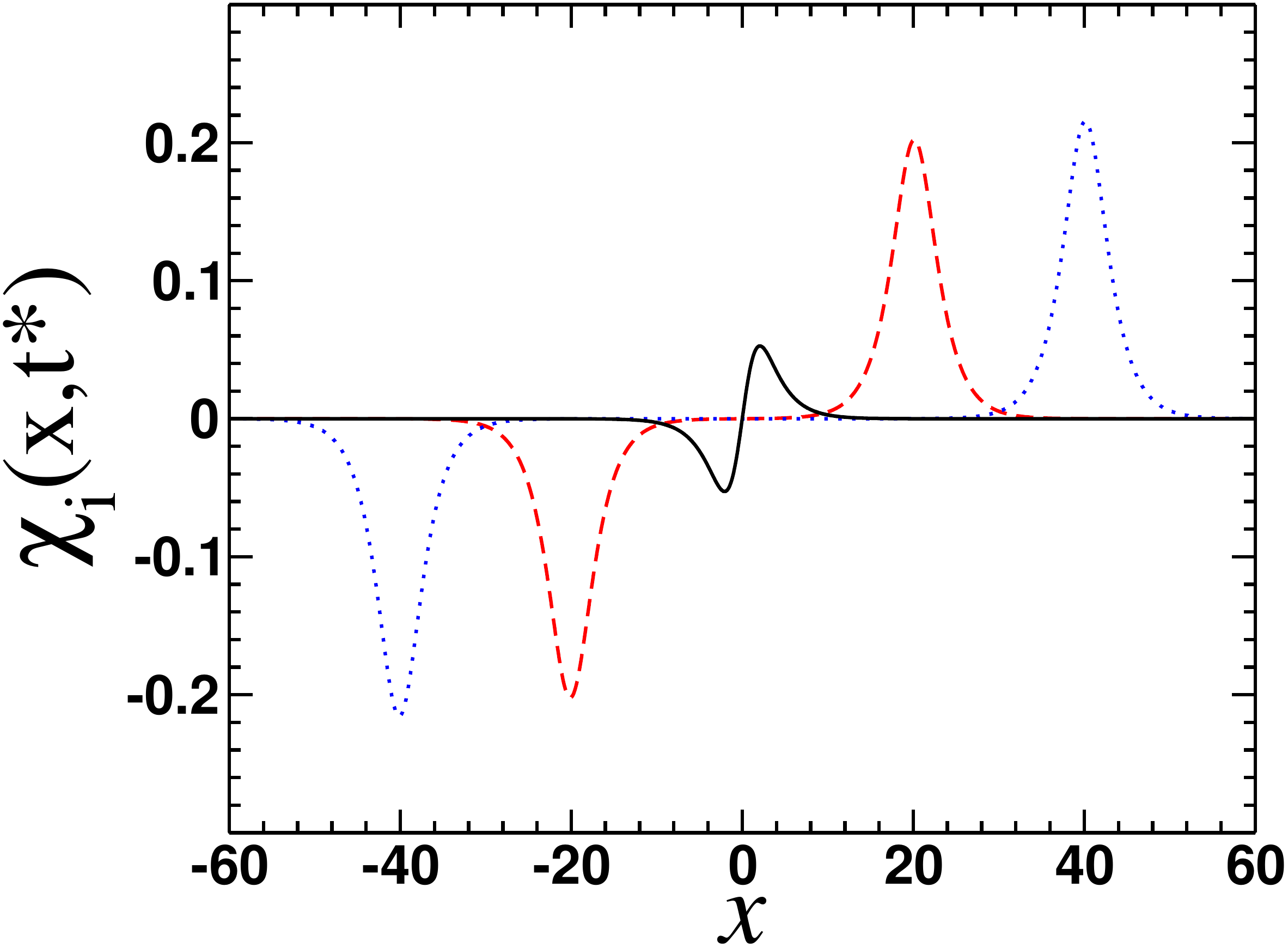}   
\end{tabular}
\end{center}
\caption{Same as Fig.\ \ref{fig7}, but for longer times: black solid lines:   
$t^{\star}=0$, red dashed lines: $t^{\star}=20$, blue dotted lines: $t^{\star}=40$. 
 }
\label{fig8} 
\end{figure}

Notice that the NLD Eq.\ 
(\ref{eq4}) with $m=0$ is invariant under the transformation $\psi(x,t) \to \psi(-x,t)$, $\chi(x,t) \to -\chi(-x,t)$ and 
$x \to -x$. Therefore, 
\begin{equation} \label{eq11a}
\psi(x,t)=\psi(-x,t), \quad \chi(x,t)=-\chi(-x,t).
\end{equation}
These symmetries are fulfilled by the numerical solutions in Fig.\ \ref{fig1},  since the initial conditions (ICs) (\ref{eq5}) with (\ref{eq8})-(\ref{eq9}) also satisfy (\ref{eq11a}). This is observed for short and long times in Figs.\  \ref{fig7} and \ref{fig8}. Figure\ \ref{fig8} shows an additional feature: for $t \gg t_s$ ($t_s$ is a transient time), i.e. for the red pulses at $t^\star=20$ and the blue pulses at $t^\star=40$, the following holds: 
\begin{eqnarray}\label{eq52}
\psi(x,t)&=& \mbox{sign}(x) \chi (x,t).  
\end{eqnarray}   
As a consequence,  
\begin{equation} \label{eq12}
\bar{\Psi} \Psi = |\psi(x,t)|^2- |\chi(x,t)|^2 \to 0.   
\end{equation} 
This means that the nonlinear term in Eq.\ (\ref{eq4}) vanishes.  

As the nonlinear term of the NLD equation approaches zero for $t \gg t_s$, the pulses which travel to the right and to the left 
with speed of light are solutions of the linearized NLD equation for $t \gg t_s$ and will be given in the next section.    
 
When we choose $\omega \ll \mu =1$, e.g. $\omega=0.1$, the initial charge density Eq.\ (\ref{eq10}) exhibits two humps. 
This facilitates the splitting into two pulses. Therefore the transient time $t_s=5$ is much shorter here than in the case $\omega=0.9$. 

Defining   
\begin{eqnarray}\label{eq18}
f(x,t) &=& \psi(x,t)+\chi(x,t), \\
\label{eq19}
h(x,t) &=& \psi(x,t)-\chi(x,t),  
\end{eqnarray}
another interesting  implication of Eq.\ (\ref{eq52}) is that 
\begin{eqnarray} \label{eq13a} 
f(x,t)&=&0, \quad x<0, \\ \label{eq13b}
h(x,t)&=&0, \quad x>0
\end{eqnarray}
 
\begin{figure}[ht!]
\begin{center}
\begin{tabular}{c} 
\includegraphics[width=7.0cm]{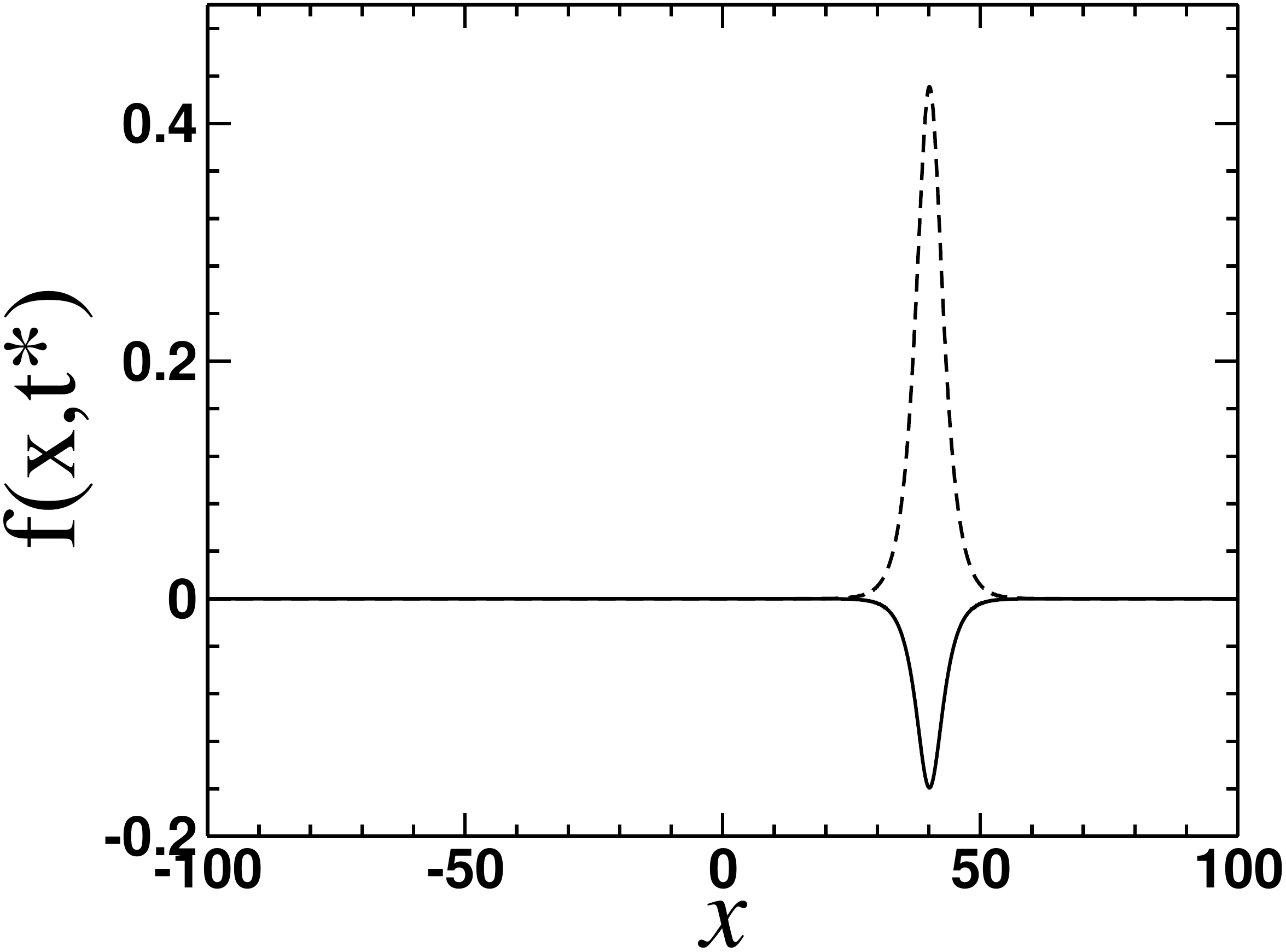} 
\end{tabular}
\end{center}
\caption{Snapshots of $f_{r}(x,t)$ (solid line) and $f_{i}(x,t)$ (dashed line) at $t=t^\star=40$. Constant potential: $V(x)=\omega=0.9$, same parameters and IC as in  Fig.\ \ref{fig1}.  
 }
\label{fig9} 
\end{figure}

As an example we present a snapshot of the real and imaginary parts of $f(x,40)$ 
for all $x$ in Fig.\ \ref{fig9}. Interestingly, the pulse in Fig.\ \ref{fig9} resembles  a \textit{semi-compacton} \cite{ahnert:2008}, because it practically vanishes for $x<0$ and on the right-hand side it vanishes exponentially.

So far we have used different potentials, but always the same initial condition, namely Eqs. (\ref{eq5}) with (\ref{eq8})-(\ref{eq9}), and for 
$t \gg t_s$ we obtained always the same scenario for the dynamics of the pulses. This result can be generalized to a class of ICs, provided that the charge density of the ICs is a localized function. 

\begin{figure}[ht!]
\begin{center}
\begin{tabular}{cc} 
\includegraphics[width=7.0cm]{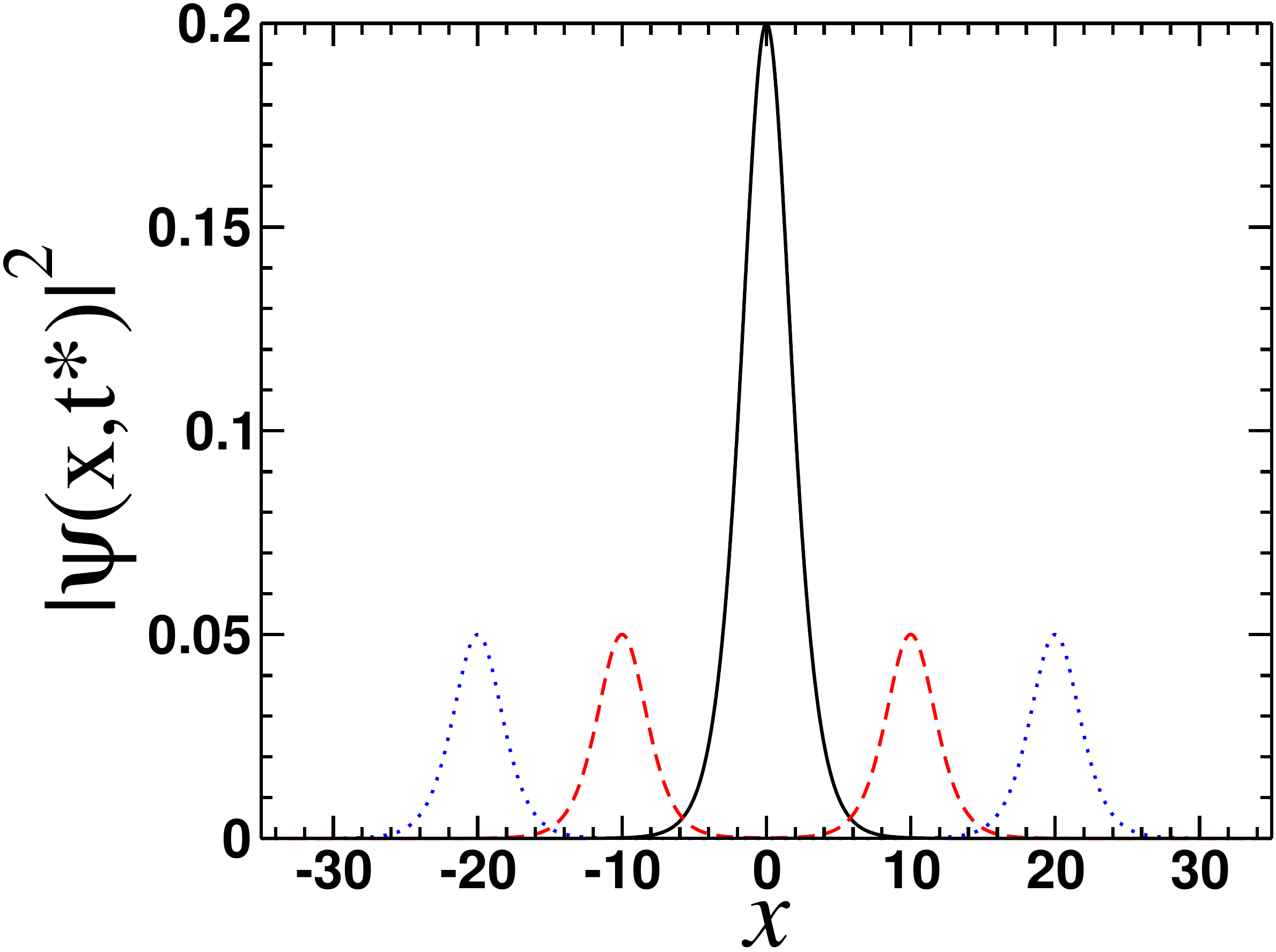} & 
\includegraphics[width=7.0cm]{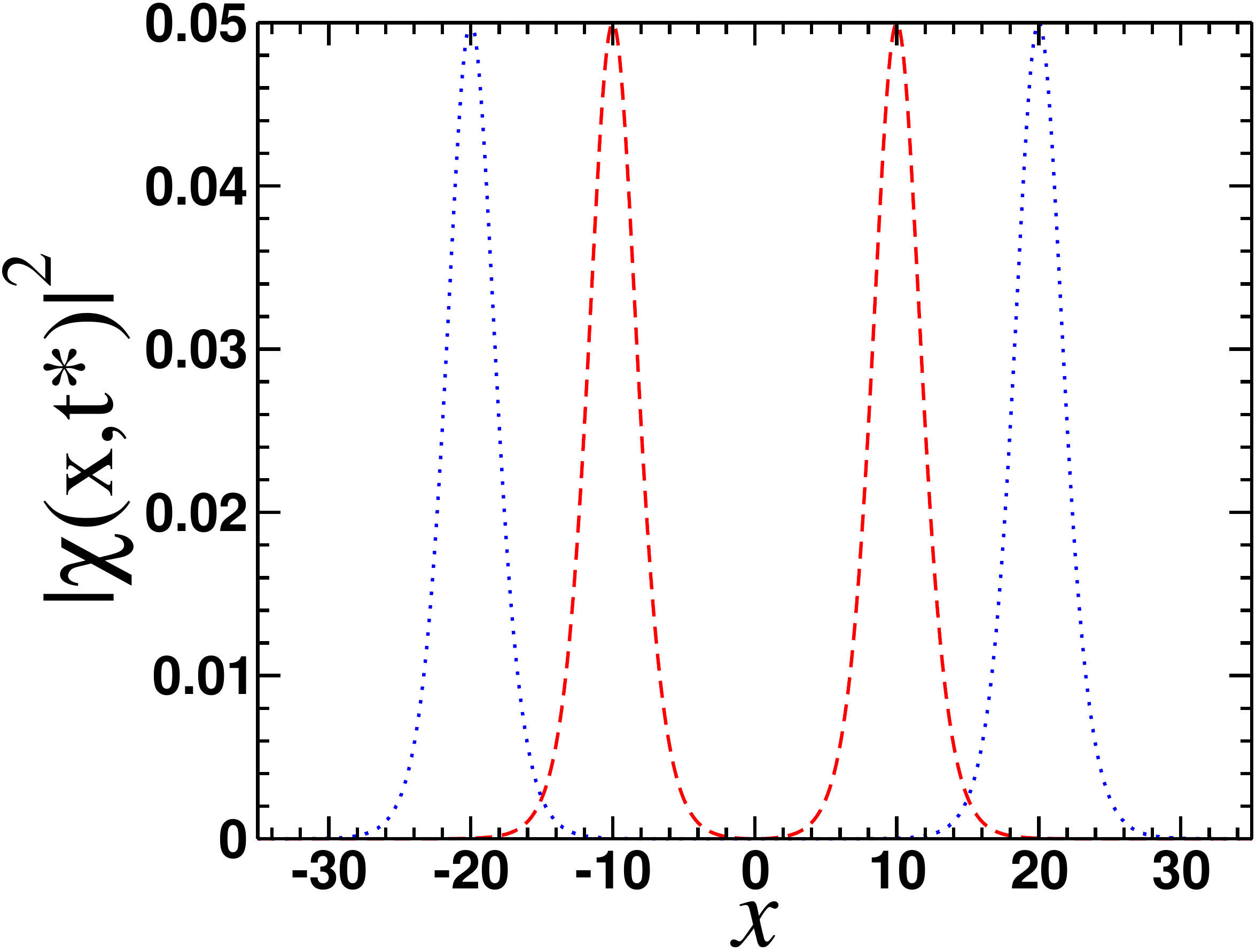}
\end{tabular}
\end{center}
\caption{Snapshots of the charge densities of the two spinor components $\psi$ and $\chi$, using the IC from the example Eq.\ (\ref{eq14}). Black solid lines: $t^{\star}=0$, red dashed lines: 
$t^{\star}=10$, and blue dotted lines:  
$t^{\star}=20$. Potential barrier $V_1(x)$, parameters: $\omega=0.9$, $\mu=1$.   
 }
\label{fig4} 
\end{figure}

As an example we present the following case: 
\begin{eqnarray}\label{eq14}
\psi(x,0)&=& 2\, a_1\, \sech\, \beta\, x, \quad 
\chi(x,0)=0,  
\end{eqnarray} 
where $a_1=\sqrt{(\mu-\omega)/2}$.  
Here Fig. \ref{fig4} shows that after splitting the two pulses again move in opposite directions with the speed of light. The charge densities $|\psi(x,t)|^2$ and $|\chi(x,t)|^2$ approach each other which means that the nonlinear term in the NLD equation  vanishes for $t \gg t_s$. In this case, $\psi(x,t)$ and $\chi(x,t)$ also possess the symmetries  (\ref{eq11a}).

\section{Exact analytical solutions of the massless NLD equation} \label{sec3}

In the previous section we have used three different potentials and also different ICs and obtained very similar results for $t \gg t_s$. Therefore, we now concentrate on the simplest case, namely a constant potential $V_2(x)=\omega$. 

 The two components of the NLD Eq.\ (\ref{eq4}) with $m=0$ satisfy  
\begin{eqnarray}\label{eq16}
i \partial_t \psi + i \partial_x \chi+g^2 \{|\psi|^2-|\chi|^2 \} \psi &=& \omega \psi \\
\label{eq17}
-i \partial_t \chi - i \partial_x \psi+g^2 \{|\psi|^2-|\chi|^2 \} \chi &=& -\omega \chi.
\end{eqnarray} 
Using the transformations  (\ref{eq18}) and (\ref{eq19}) 
we obtain 
\begin{eqnarray}\label{eq20}
\partial_t f + \partial_x f -i \frac{g^2}{2} \{f h^\star+f^\star h \} h &=& -i \omega f, \\
\label{eq21}
\partial_t h - \partial_x h -i \frac{g^2}{2} \{f h^\star+f^\star h \} f &=& -i \omega h. 
\end{eqnarray} 
 First, we seek localized solutions of the \textit{linearized} Eqs.\ (\ref{eq20})-(\ref{eq21})
\begin{eqnarray}\label{eq22}
\partial_t f + \partial_x f+ i \omega f &=&0, \\
\label{eq23}
\partial_t h - \partial_x h+ i \omega h &=& 0, 
\end{eqnarray} 
with ICs 
\begin{eqnarray}\label{eq24}
f(x,0)= \rho_1(x) e^{i \theta_{1}(x)}, \quad 
h(x,0)= \rho_2(x) e^{i \theta_{2}(x)}, 
\end{eqnarray} 
where the amplitudes $\rho_1$ and $\rho_2$ are real functions, which go to zero for $x \to \pm \infty$, and the phases 
$\theta_1$ and $\theta_2$ are also real functions. 
The equations (\ref{eq22})-(\ref{eq23}) are decoupled and have the solutions
\begin{eqnarray}\label{eq26}
f(x,t)= e^{-i \omega t} \rho_1(x-t) e^{i \theta_{1}(x-t)}, \quad 
h(x,t)= e^{-i \omega t} \rho_2(x+t) e^{i \theta_{2}(x+t)}. 
\end{eqnarray}
Substituting 
Eqs.\ (\ref{eq26}) in Eqs.\ (\ref{eq18})-(\ref{eq19}) we obtain 
\begin{eqnarray}\label{eq28}
\psi(x,t)&=& \frac{1}{2}  e^{-i \omega t} \left[\rho_1(x-t) e^{i \theta_{1}(x-t)}+ \rho_2(x+t)  e^{i \theta_{2}(x+t)}\right], \\
\label{eq29}
\chi(x,t)&=& \frac{1}{2}  e^{-i \omega t} \left[\rho_1(x-t) e^{i \theta_{1}(x-t)}- \rho_2(x+t)  e^{i \theta_{2}(x+t)}\right].
\end{eqnarray}
This is the solution of the linearized massless NLD equation. It is composed of two localized waves with the shapes $\rho_1(x)$ 
and $\rho_2(x)$, which travel in opposite directions with the speed of light, namely unity. 

 This is very similar to the numerical solutions of the full NLD equation for $t \gg t_s$, which we obtained in Sec.\ \ref{sec2}. 
 For this time regime the nonlinear term in Eq.\ (\ref{eq4}) vanished because $\bar{\Psi} \Psi=|\psi(x,t)|^2-|\chi(x,t)|^2 \to 0$. 
 
 As we want to find exact analytical solutions of the \textit{full} NLD equation we require that the nonlinear terms in Eqs.\ (\ref{eq16})-(\ref{eq17}) and (\ref{eq20})-(\ref{eq21}) 
 vanish for all times
\begin{equation}\label{eq30}
|\psi|^2-|\chi|^2=\frac{1}{2} [f h^\star + f^\star h]=0. 
\end{equation}
When this condition is fulfilled, Eqs.\  (\ref{eq28})-(\ref{eq29}) are exact solutions of the full NLD equation for all times. The condition, Eq.\ (\ref{eq30}), explicitly reads
\begin{equation}\label{eq31}
\rho_1(x-t) \rho_2(x+t) \cos \Theta(x,t)=0, 
\end{equation}
 with $\Theta(x,t)= \theta_{2}(x+t)-\theta_{1}(x-t)$, and is fulfilled in three cases:
\begin{equation}\label{eq33}
\rho_1(x-t)=0, 
\end{equation}
or 
\begin{equation}\label{eq33a}
\rho_2(x+t)=0, 
\end{equation}
or
 \begin{equation}
 \label{eq32}
\theta_{2}(x+t)-\theta_{1}(x-t)=\pm \frac{\pi}{2}.  
\end{equation}
For $\rho_1 \equiv 0$, here $f(x,t)=0$ and $\psi(x,t)=-\chi(x,t)=\frac{1}{2} e^{i [\theta_2(x+t)-\omega t]} \rho_2(x+t)$. In this case a pulse with the shape $\rho_2(x)$ travels to the left with the speed of light. Similarly, for $\rho_2 \equiv 0$ a pulse travels to the right. 
 
Let us consider Eq.\ (\ref{eq32}). As it must be fulfilled for all $x$ and $t$, the only solution is that $\theta_1$ and $\theta_2$ are constants. 
Thus, choosing the minus sign in the condition (\ref{eq32}),  the other exact solution of the massless NLD equation finally reads 
\begin{eqnarray}\label{eq38}
\psi(x,t)&=& \frac{1}{2}  e^{i (\theta_1-\omega t)} \left[\rho_1(x-t) -i \rho_2(x+t)\right], \\
\label{eq39}
\chi(x,t)&=& \frac{1}{2} e^{i (\theta_1-\omega t)} \left[\rho_1(x-t) +i \rho_2(x+t)\right].  
\end{eqnarray}
This solution contains an arbitrary constant $\theta_1$ 
and  arbitrary, localized functions $\rho_1(x,t)$ and $\rho_2(x,t)$, whereas the solution (\ref{eq28})-(\ref{eq29}) of the linearized massless NLD equation contained four arbitrary functions $\theta_1(x,t)$, 
$\theta_2(x,t)$, $\rho_1(x,t)$ and $\rho_2(x,t)$. Considering instead the plus sign  in  (\ref{eq32}) is equivalent to interchanging   
$\psi(x,t)$ with $\chi(x,t)$. Clearly, this interchange leaves invariant  Eqs.\ (\ref{eq16})-(\ref{eq17}). 

To proceed it is shown that if the IC together with 
the massless NLD equation satisfy further symmetries, 
the solution given by  Eqs.\ (\ref{eq38})-(\ref{eq39}) is  simplified.  Indeed, Eqs.\  
(\ref{eq16})-(\ref{eq17})  are invariant under the transformation $\psi(x,t) \to -i \chi(-x,t)$, 
and $x \to -x$. This means that 
\begin{equation}\label{eq34}
\psi(x,t) = -i \chi(-x,t).
\end{equation}
The exact solutions Eqs.\ (\ref{eq38})-(\ref{eq39}) 
satisfy the symmetry (\ref{eq34}) only when 
\begin{eqnarray}
\label{eq37}
 \rho_1(x-t) & = & \rho_2(-x+t), \quad \rho_1(x+t)  =  \rho_2(-x-t). 
\end{eqnarray}
In this case, the exact solution of the massless NLD equation reads
\begin{eqnarray}\label{eq38a}
\psi(x,t)&=& \frac{1}{2}  e^{i (\theta_1-\omega t)} \left[\rho_1(x-t) -i \rho_1(-x-t)\right], \\
\label{eq39b}
\chi(x,t)&=& \frac{1}{2} e^{i (\theta_1-\omega t)} \left[\rho_1(x-t) +i \rho_1(-x-t)\right].
\end{eqnarray}

\begin{figure}[ht!]
\begin{center}
\begin{tabular}{cc}
\includegraphics[width=7.0cm]{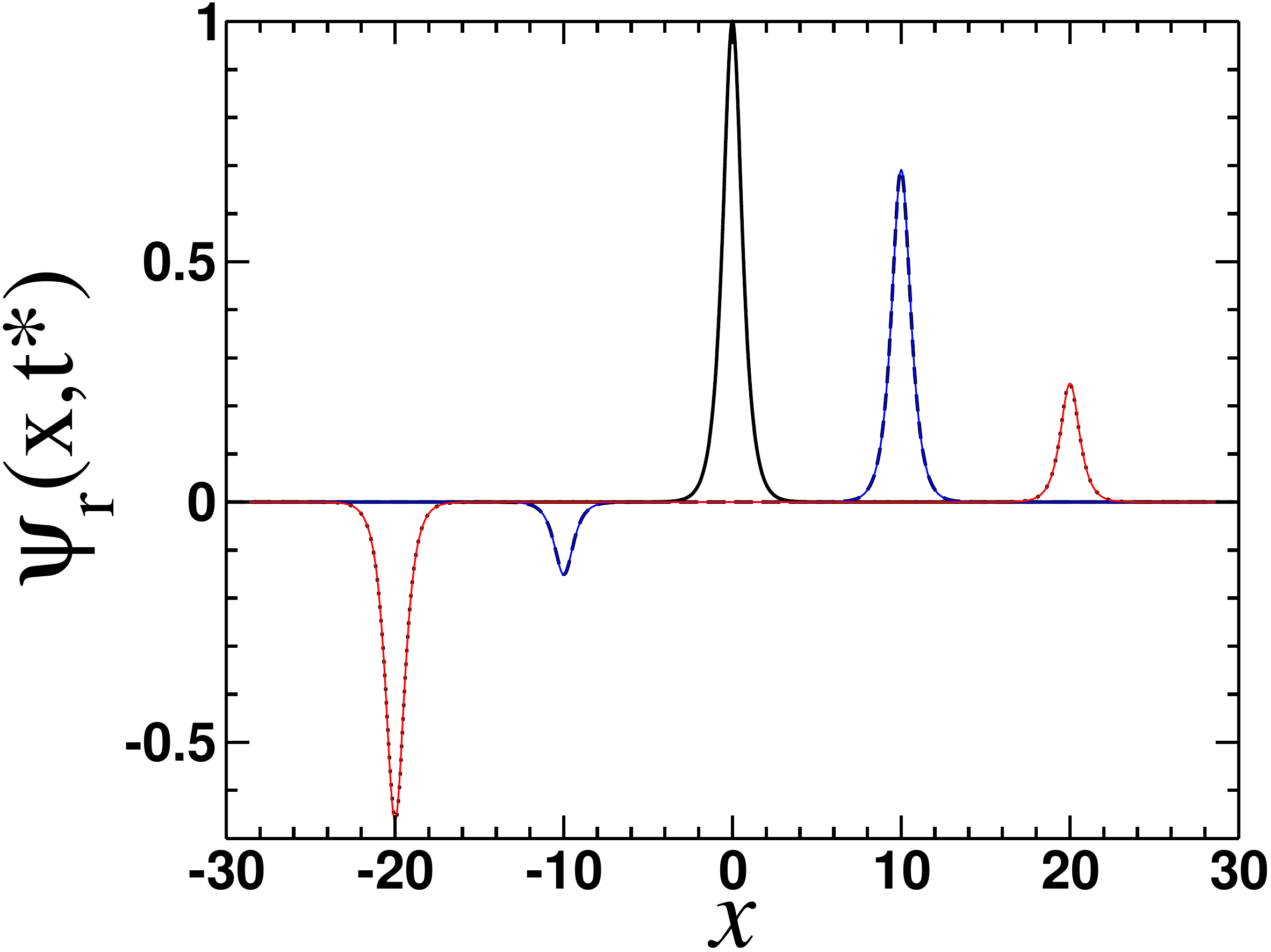} & 
  \includegraphics[width=7.0cm]{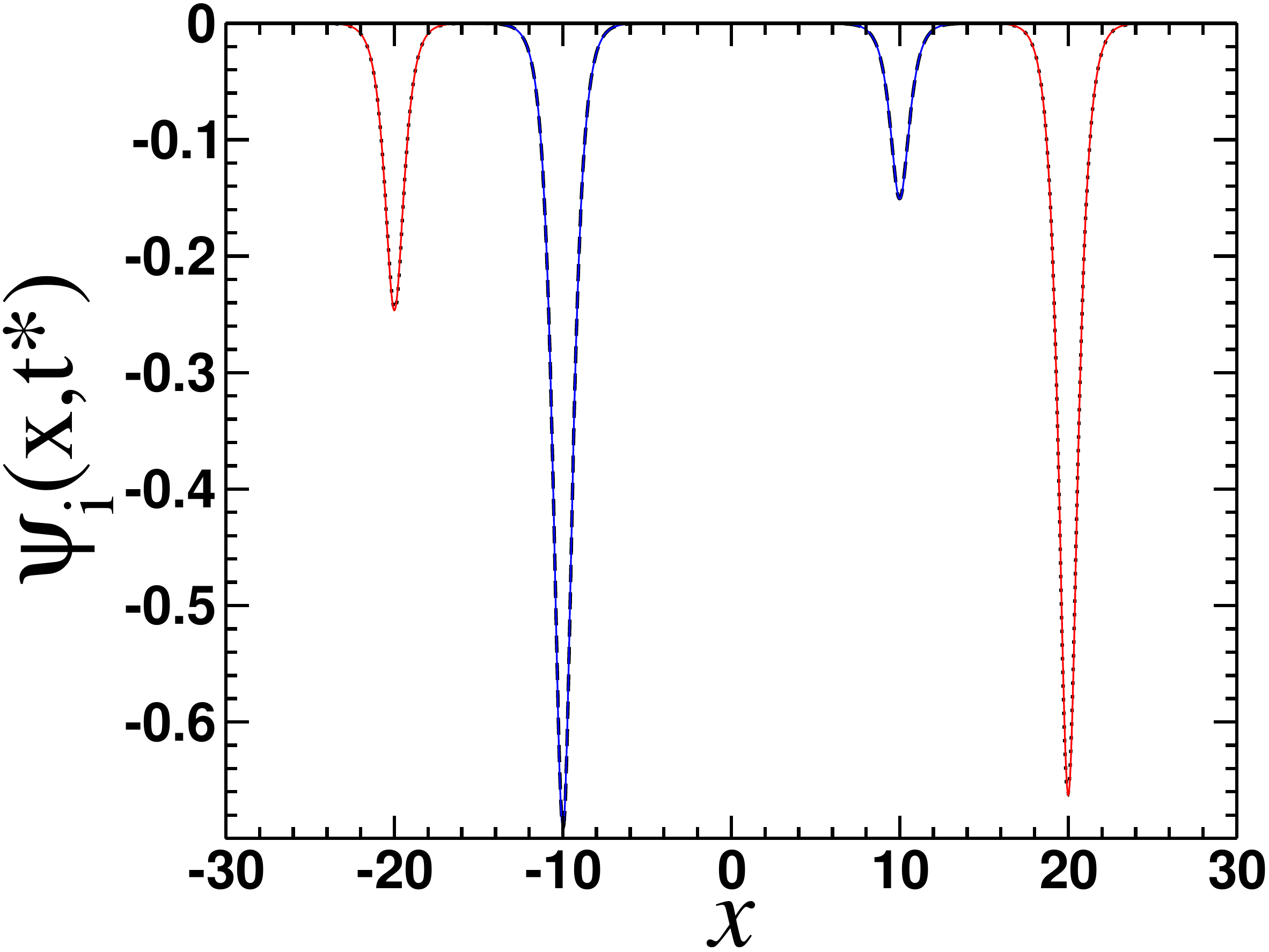} \\
 \includegraphics[width=7.0cm]{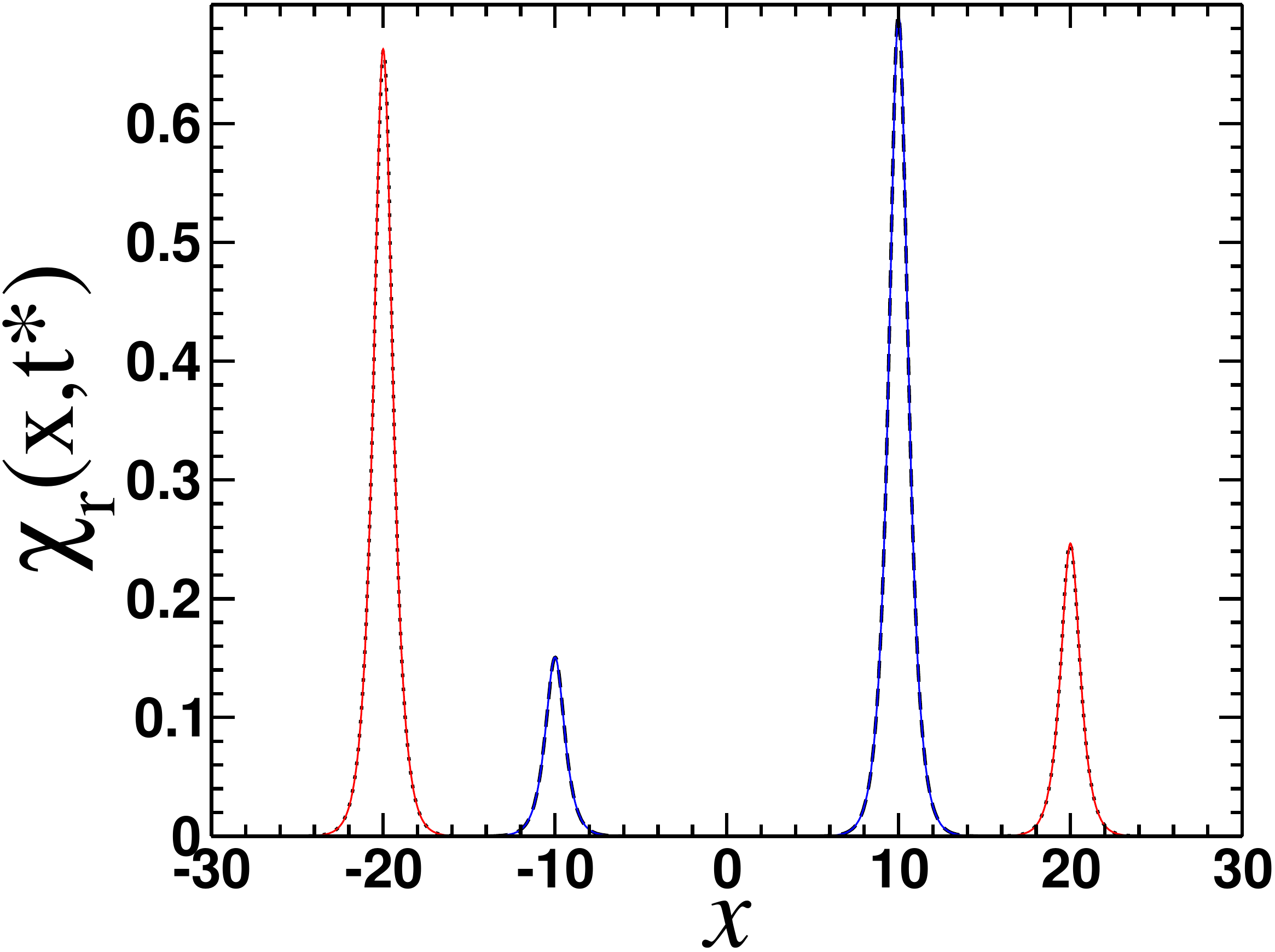} &   
 \includegraphics[width=7.0cm]{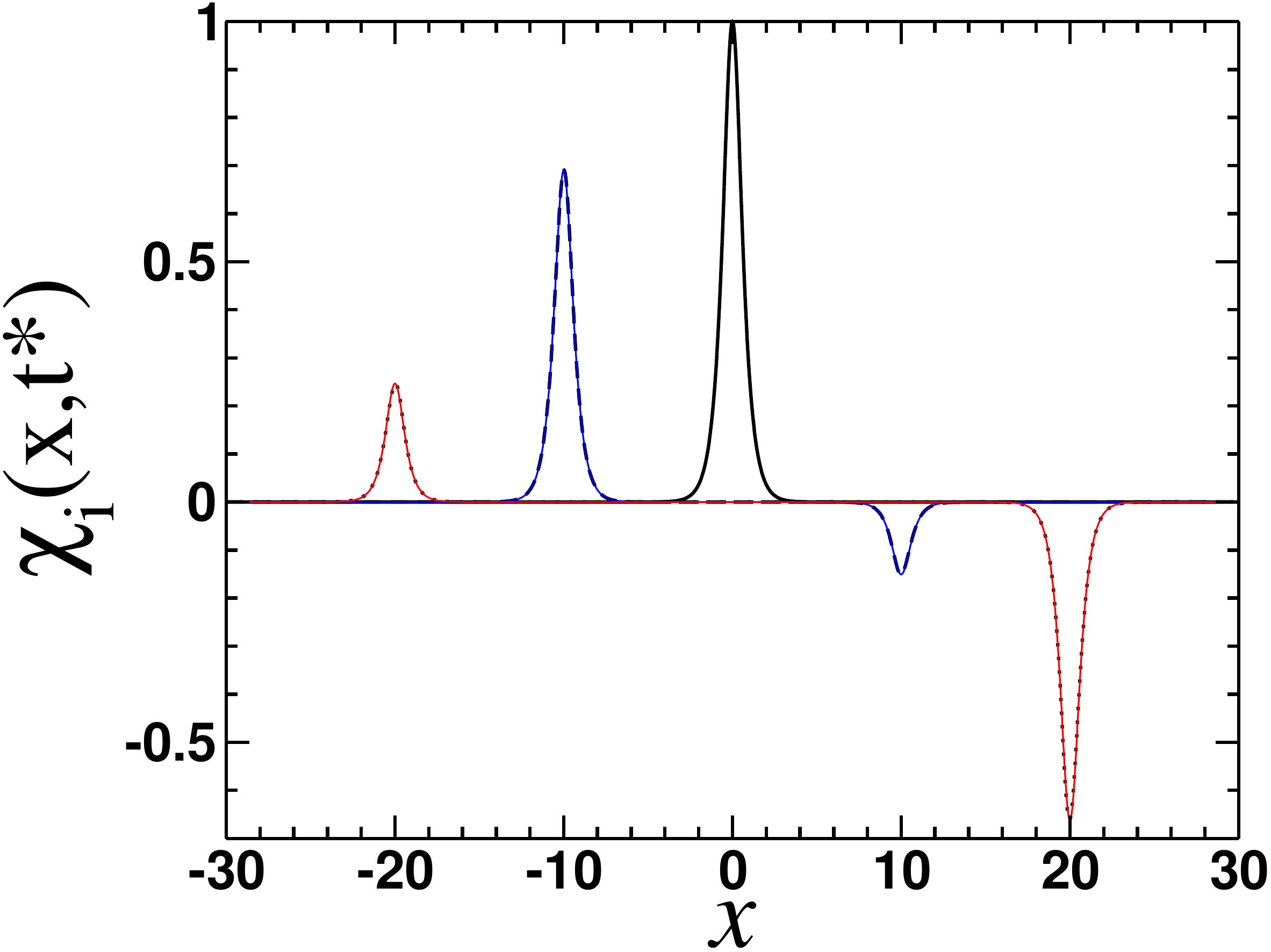}     
\end{tabular}
\end{center}
\caption{  
Real and imaginary parts of the spinor components $\psi$ and $\chi$, using the ICs given by Eq.\ (\ref{eq44}).  
Simulation: solid lines: $t^\star=0$, dashed lines: $t^\star=10$  and dotted lines: $t^\star=20$.  
Blue and red  lines, exact  solutions at $t^\star=10,20$, respectively, are overimposed with the dashed and dotted lines.
Parameters: $m=0$, $g=1$, $\omega=0.1$, $a=1$ and $b=2$.}
\label{fig5} 
\end{figure}

\begin{figure}[ht!]
\begin{center}
\begin{tabular}{cc}
\includegraphics[width=7.0cm]{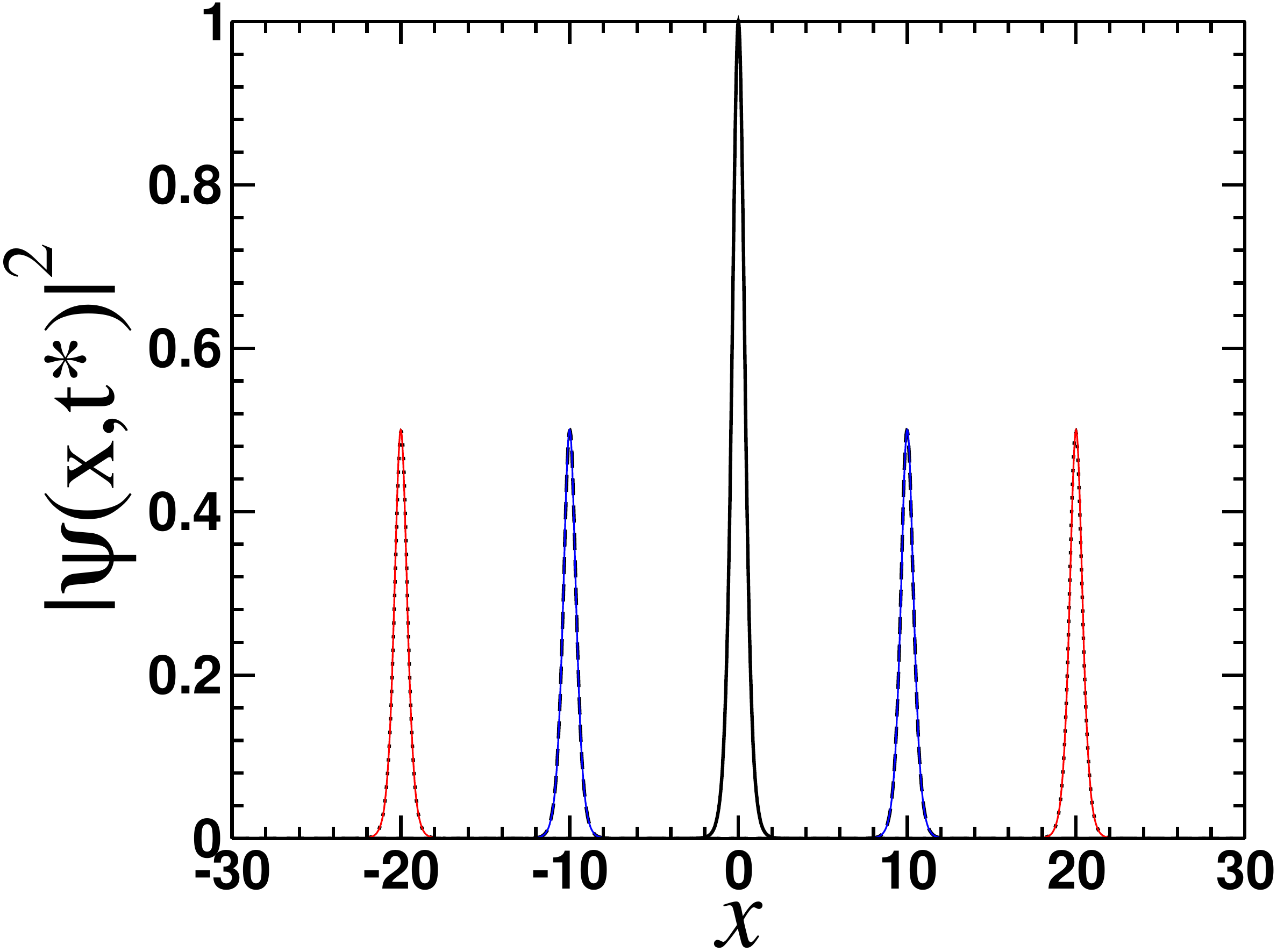} & 
  \includegraphics[width=7.0cm]{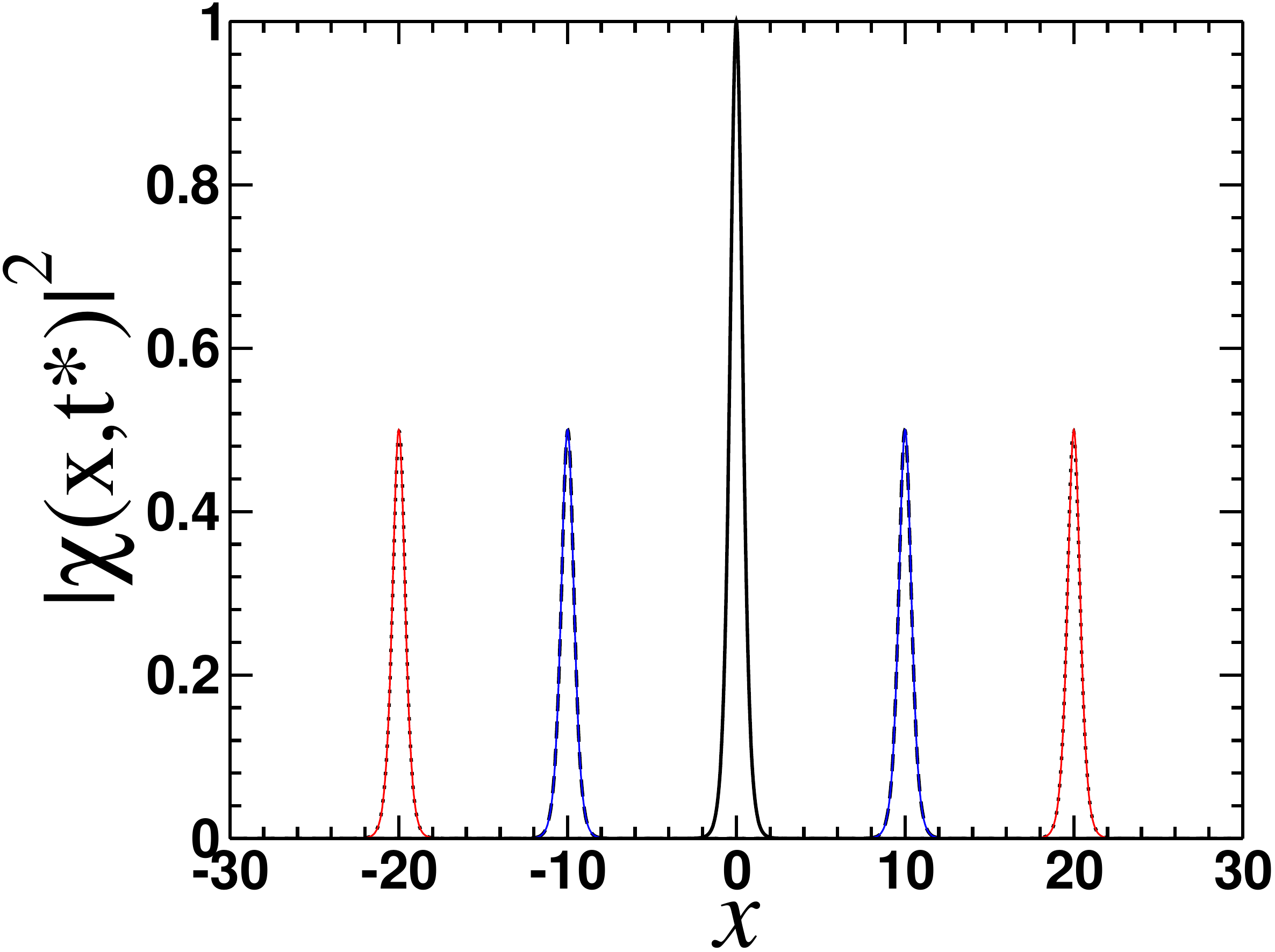} \\
 \includegraphics[width=7.0cm]{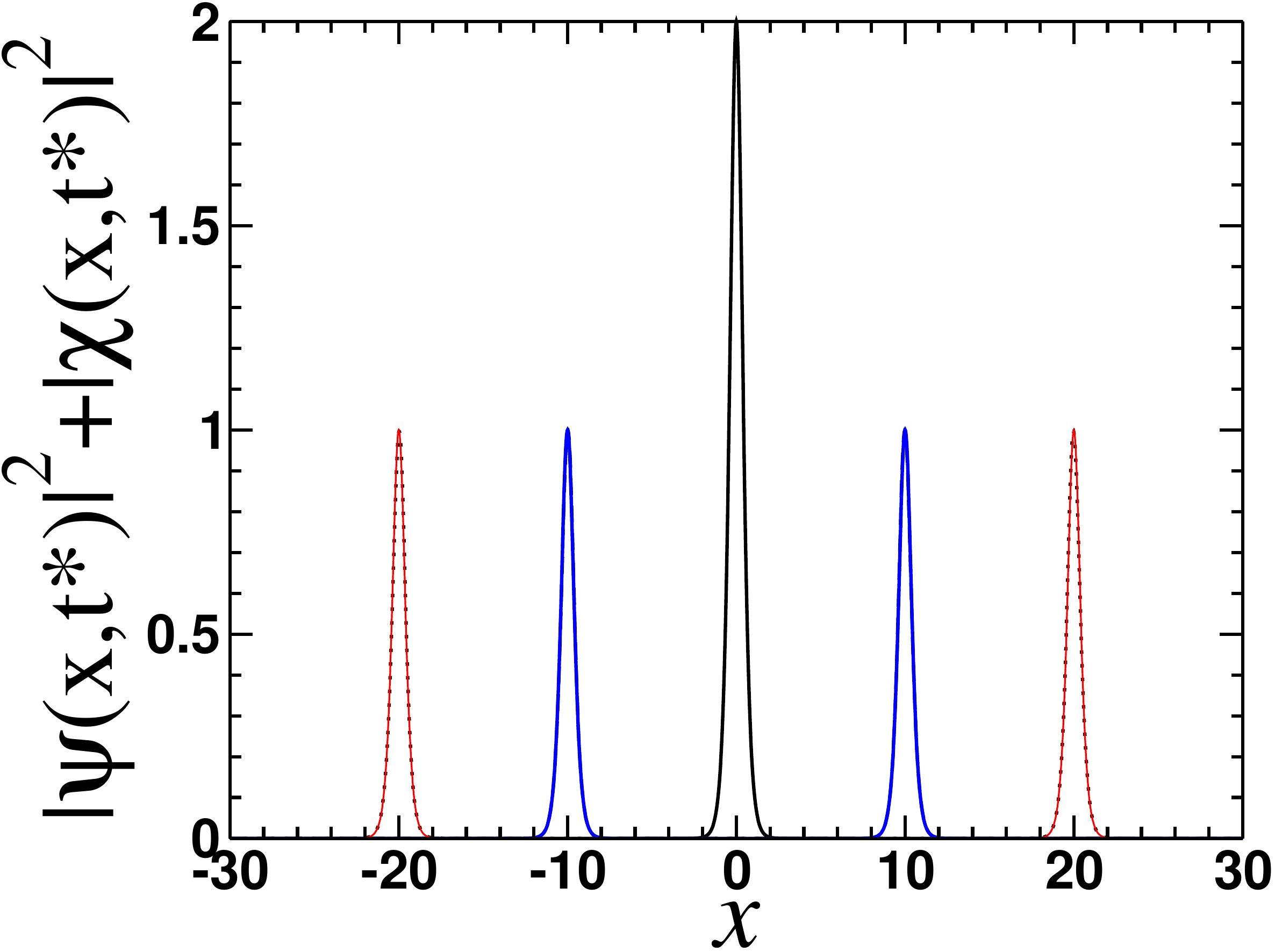} & 
 \includegraphics[width=7.0cm]{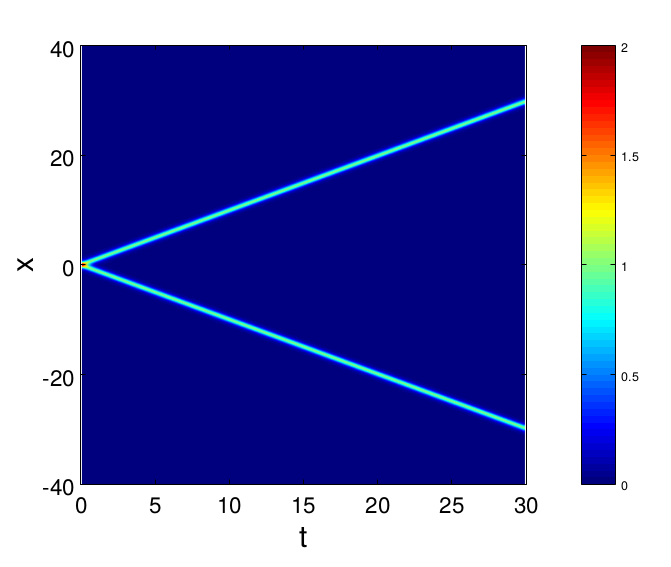}      
\end{tabular}
\end{center}
\caption{  
Charge densities of spinor components. 
Simulation: solid lines: $t^\star=0$, dashed lines: $t^\star=10$  and dotted lines: $t^\star=20$.  
Blue and red  lines, exact  solutions at $t^\star=10,20$, respectively, are overimposed with the dashed and dotted lines.   
  Right lower panel: contour plot of the charge density in the left lower panel. Same parameters and ICs as in Fig.\ \ref{fig5}.
 }
\label{fig6} 
\end{figure}

 This has an important consequence: In Eqs.\ (\ref{eq38})-(\ref{eq39}) the right and the left running pulses generally have different shapes $\rho_1(x)$ and $\rho_2(x)$, whereas now 
$\rho_1(x)$ and $\rho_2(x)$ satisfy Eq.\ (\ref{eq37}). A simple example is to 
choose 
 \begin{equation}\label{eq44}
\psi(x,0)=  a \, \mbox{sech}[b x], \quad 
\chi(x,0)= i a \, \mbox{sech}[b x],   
\end{equation}
which fulfills the symmetry in Eq.\ (\ref{eq34}), where $a$ and $b$ are constants. Notice that in this case, 
$\rho_1(x,0)=\rho_2(x,0)=\rho(x,0)=\sqrt{2} a \, \mbox{sech} [b x]$,  
 $\theta_1=\pi/4$ and $\theta_2=-\pi/4$. 
 The exact solution (\ref{eq38})-(\ref{eq39}) now becomes 
 \begin{eqnarray}\label{eq42}
\psi(x,t)&=& \frac{1+i}{2} a \, e^{-i \omega t} \{\mbox{sech}[b(x-t)]- i \mbox{sech}[b(x+t)]\},\\
\label{eq43}
\chi(x,t)&=& \frac{1+i}{2} a \, e^{-i \omega t} \{\mbox{sech}[b(x-t)]+ i \mbox{sech}[b(x+t)]\}.  
\end{eqnarray}
 Figures.\ \ref{fig5} and \ref{fig6} show the solutions represented by Eqs.\ (\ref{eq42})-(\ref{eq43}). These solutions serve as a test of our simulations, i.e. the numerical solution of the massless NLD equation, using the ICs, Eqs.\  (\ref{eq44}). The simulation also confirms that the charge and the energy are the constants $Q=4 a^2/|b|=2$ and $E = \omega Q=0.2$, respectively, predicted by the theory. Moreover, the momentum $P$ is zero. 
 
Other symmetries, as for instance, the ones represented by Eq.\ (\ref{eq11a}) drastically reduce Eqs.\ (\ref{eq38})-(\ref{eq39}) to the trivial solution $\rho_1(x,t)=\rho_2(x,t)=0$, i.e. $\psi(x,t)=\chi(x,t)=0$.  
 
\section{Conclusions}

In this paper we have investigated a notable property of the massless NLD equation in the presence of an external field-- the fact that initial pulses centered
at the origin of the external field break into two pulses traveling in opposite directions, which after a short time become solutions of the {\it massless} linear
Dirac equation.  We have presented numerical simulations starting with various pulse initial conditions for the case of three different
external potentials: a potential barrier, a potential well and a constant potential.  By considering  exact solutions of the massless NLD equation in the presence of a constant  potential which are {\it also} solutions of the linear Dirac equation we were able to gain insight into why an initial
pulse at time zero becomes two solutions of the linear Dirac equation moving in opposite directions at later times.  The latter exact solution also served as a 
test for our numerical procedure.

\section{Appendix: Conservation laws}

The NLD Eq.\ (\ref{eq4}) can be derived in a standard fashion from the Lagrangian density
\bq \label{A1}
\mathcal{L} =  \left(\frac{i}{2}\right) [\bPsi \gamma^{\mu} \partial_{\mu} \Psi 
-\partial_{\mu} \bPsi \gamma^{\mu} \Psi] - m \bPsi \Psi 
+ \frac{g^2}{2} (\bPsi \Psi)^{2} - \bPsi \gamma^{0} V(x) \Psi \>.
\eq
Multiplying Eq. (\ref{eq4}) to the left by $\bPsi$, multiplying the adjoint NLD equation to the right by $\Psi$ and summing up, we obtain the continuity equation
\bq \label{A2}
\frac{\partial}{\partial t} (\bPsi \gamma^{0} \Psi)+\frac{\partial}{\partial x} (\bPsi \gamma^{1} \Psi)=0,
\eq
where $\bPsi \gamma^{0} \Psi=|\psi|^2+|\chi|^2$ is the density of the charge. Integrating over $x$ and assuming that 
$\Psi$ vanishes at infinity we obtain that the charge is conserved, i.e. 
\bq \label{A3}
\frac{dQ}{dt}=\frac{\partial}{\partial t} \int dx (\bPsi \gamma^{0} \Psi)=0. 
\eq

Multiplying Eq. (\ref{eq4}) to the left by $\bPsi_{x}$, multiplying the adjoint NLD equation to the right by $\Psi_x$ and taking the difference between these two expressions, we obtain the continuity equation
\bq \label{A4}
\frac{\partial T^{01}}{\partial t} +\frac{\partial T^{11}}{\partial x} = - \bPsi \gamma^{0} 
\frac{\partial V}{\partial x} \Psi,
\eq
where the density of the momentum is 
\bq \label{A5}
T^{01}=\frac{i}{2} (\bPsi_x \gamma^{0} \Psi-\bPsi \gamma^{0} \Psi_x),
\eq
\bq \label{A6}
T^{11} = -m \bPsi \Psi+ \frac{g^2}{2} (\bPsi \Psi)^2+\frac{i}{2} [\bPsi \gamma^{0} \Psi_t-\bPsi_t \gamma^{0} \Psi]- \bPsi \gamma^{0} V(x) \Psi . 
\eq
Integrating over $x$ and assuming that 
$T^{11}(-\infty)-T^{11}(\infty)=0$ we obtain  
\bq \label{A7}
\frac{dP}{dt}=-\int dx \bPsi \gamma^{0} 
\frac{\partial V}{\partial x} \Psi, 
\eq  
where the momentum is 
\bq \label{A8}
P=\int dx T^{01}.
\eq
In the case of $V(x)=V_2(x)=\omega$, the momentum is conserved. Moreover, if the potential is symmetric, i.e. $V(x)=V(-x)$ and the spinor field satisfies either the 
symmetry in Eq.\ (\ref{eq11a}) or in Eq.\ (\ref{eq34}), it can be shown that $P$ is also conserved.

Multiplying Eq. (\ref{eq4}) to the left by $\bPsi_{t}$, multiplying the adjoint NLD equation to the right by 
$\Psi_t$ and taking the difference between these two expressions, we obtain the continuity equation
\bq \label{A9}
\frac{\partial T^{00}}{\partial t} +\frac{\partial T^{10}}{\partial x} = \bPsi \gamma^{0} 
\frac{\partial V}{\partial t} \Psi=0,
\eq
where  
\bq \label{A10}
T^{10}=\frac{i}{2} (\bPsi \gamma^{1} \Psi_t-\bPsi_t \gamma^{1} \Psi) , 
\eq
and the density of the energy is 
\bq \label{A11}
T^{00}=-\frac{i}{2} (\bPsi \gamma^{1} \Psi_x-\bPsi_x \gamma^{1} \Psi)+m \bPsi \Psi-\frac{g^2}{2} (\bPsi \Psi)^2+
\bPsi \gamma^0 V(x) \Psi.
\eq
Integrating over $x$ Eq.\ (\ref{A9}), the energy, $E=\int dx \, T^{00}$, is conserved if $T^{10}(-\infty)-T^{10}(+\infty)=0$.

\begin{acknowledgments}
We acknowledge financial support from the Ministerio de Econom\'ia y Competitividad of Spain through Grant No. FIS2014-54497-P
(N.R.Q. and F.G.M.), from the Junta de Andaluc\'ia through Grant No.
P11-FQM-7276 (N.R.Q.) and from  the University of Seville through the Plan Propio (F.G.M.). F.G.M. acknowledges financial support and hospitality of the Theoretical Division and Center for Nonlinear Studies at Los Alamos national Laboratory.  This work was supported in part by the U.S. Department of Energy. 
\end{acknowledgments}

\bibstyle{revtex}

\bibliography{dirac2}

\end{document}